\documentclass[a4paper,10pt]{article}
\usepackage{jcappub}

\usepackage[figuresright]{rotating}
\usepackage{amsmath}
\usepackage{amsfonts}
\usepackage{amssymb}
\usepackage{bbold}
\usepackage{xfrac}
\usepackage{mathrsfs}
\usepackage{bm}
\usepackage{listings}
\usepackage{natbib}
\usepackage[utf8]{inputenc}
\usepackage{graphicx}
\usepackage{epsfig}
\usepackage{float}
\usepackage{subcaption}
\graphicspath{{./Pics/}}
\usepackage{cancel}
\usepackage{amsthm}
\usepackage{hyperref}
\usepackage{titlesec}
\usepackage{fancyvrb}
\usepackage{dsfont}
\usepackage{scrextend}
\usepackage{enumitem}
\usepackage{framed}
\usepackage{empheq}
\usepackage{booktabs}
\usepackage[listings]{tcolorbox}
\usepackage{lipsum}

\def \pa{\partial}

\def \tcb{\textcolor{blue}}

\begin{document}

\title{Mass varying neutrinos with different quintessence potentials}

\author[a,b,c]{Sayan~Mandal}
\author[d]{Gennady~Y.~Chitov}
\author[c]{Olga~Avsajanishvili}
\author[b]{Bijit~Singha}
\author[b,c,d]{Tina~Kahniashvili}

\affiliation[a]{Physics and Astronomy Department, Stony Brook University, 100 Nicolls Road, Stony Brook, New York 11794, USA}

\affiliation[b]{McWilliams Center for Cosmology and Department of Physics, Carnegie Mellon University, 5000 Forbes Avenue, Pittsburgh, Pennsylvania 15213, USA}

\affiliation[c]{Faculty of Natural Sciences and Medicine, Ilia State University, 3-5 Cholokashvili Street, 0194 Tbilisi, Georgia}

\affiliation[d]{Department of Physics, Laurentian University, Ramsey Lake Road, Sudbury, Ontario, P3E 2C6, Canada}

\emailAdd{sayan.mandal@stonybrook.edu}

\date{\today}
\abstract{
The mass-varying neutrino scenario is analyzed for three trial quintessence potentials (Ferreira-Joyce, inverse exponential,
and thawing oscillating).
The neutrino mass is generated via Yukawa coupling to the scalar field which represents dark energy.
The inverse exponential and oscillating potentials are shown to successfully generate the neutrino masses in the range  $m \sim 10^{-2}-10^{-3}~$eV
and to yield the current dark energy density in the regime of the late-time acceleration of the Universe.
Depending on the choice of potentials, the acceleration could occur in two different regimes:
(1) the regime of instability, and (2) the stable regime. The first regime of instability is after the Universe underwent a first-order transition and is rolling toward the new stable vacuum. The imaginary sound velocity $c^2_s < 0$ in this regime implies growing fluctuations of the neutrino density (clustering). In the second regime, the Universe smoothly changes its stable states via a continuous transition. Since $c^2_s > 0$, the neutrino density is stable.
For all cases the predicted late-time acceleration of the Universe is asymptotically very close to that of the $\Lambda$CDM model.
Further extensions of the theory to modify the neutrino sector of the Standard Model and to incorporate inflation are also discussed.
It is also shown that in the stable regimes where the neutrino mass is given by the minimum of the thermodynamic potential, the tree-level dynamics of the scalar field is robust with respect to one-loop bosonic and fermionic corrections to the potential.
}

\maketitle

\section{Introduction}
\label{intro}

Understanding the origin of neutrino masses is one of the biggest questions in modern high energy physics and cosmology. Various experiments have conclusively established the fact that neutrinos have a small mass (see Sec. 14.1. of Ref.~\cite{Tanabashi:2018oca}). Cosmological observations have put a stringent upper bound \cite{Aghanim:2018eyx} on the total neutrino mass of about $0.12\,\mathrm{eV}$,
while the tightest bounds of $1.1\,\mathrm{eV}$ from kinematic measurements have recently been reported by the KATRIN collaboration \cite{Aker:2019uuj}. On the other hand, lower bounds of $0.06\,\mathrm{eV}$ come from measurement of the squared mass difference in oscillation experiments (see Table 2.1 and Sec 14 of Ref.~\cite{Tanabashi:2018oca}). It is widely known that the origin of neutrino masses will hold a clue to our understanding of microscopic physics beyond the Standard Model of particle physics (see Ref.~\cite{Mohapatra:2006gs} for a pioneering reference; see also Ref.~\cite{Kang:2019xuq,Abazajian:2017tcc,King:2017guk}, among others). Neutrinos are massless in the Standard Model of particle physics since there are no right handed neutrinos, and thus a large number of beyond-the-Standard-Model theories have been proposed to explain the origin of neutrino masses. These include the seesaw mechanisms, $R$-parity violating supersymmetry, and theories based on extra dimensions, among others; a comprehensive review is provided in Ref.~\cite{King:2015aea}.

Another outstanding problem in modern cosmology is trying to understand the origin and properties of dark energy (DE), the hypothetical substance generally believed to be causing the observed accelerated expansion of the Universe. Current observations suggest that about 68\% of the Universe is dominated by this mysterious form of energy \cite{Aghanim:2018eyx}. Several authors propose that the accelerated expansion can be explained by a modification of the general theory of relativity (see Refs.  \cite{Capozziello:2003tk,Nojiri:2003ft,Carroll:2004de} for some early work, \cite{Joyce:2016vqv} for a recent review, and \cite{Slosar:2019flp} for updates on some developments). Within the framework of general relativity, there are two most popular candidates for DE, namely the cosmological constant (see Refs.~\cite{Peebles:2002gy,Caldwell:2009ix,Silvestri:2009hh} for reviews) and the scalar field DE, known also as dynamical DE (comprising of quintessence \cite{Ratra:1987rm,Peebles:1987ek,Wetterich:1987fm,Caldwell:2005tm} and phantom DE \cite{Caldwell:1999ew,Scherrer:2008be,Dutta:2009dr,Ludwick:2017tox}; see Ref.~\cite{Amendola:2015ksp} for a more detailed discussion on the classification and observational tests). The dynamics in the latter case is assumed to be similar to the scalar field dynamics purported to cause inflationary expansion of the early Universe (see Refs.~\cite{Guth:1980zm,Starobinsky:1982ee,Linde:1981mu} for the original works and \cite{Baumann:2008bn,Martin:2013tda,Senatore:2016aui} for reviews). In a recent work \cite{Avsajanishvili:2017zoj}, the authors have obtained observational constraints on various DE potentials.

A related puzzle is the \textit{coincidence problem} -- at the present epoch, the energy densities of dark matter (DM) and DE are of comparable magnitudes, and these are again comparable to that of cosmological neutrinos within a few orders of magnitude. A lot of work has been done (see e.g., Refs.~\cite{Farrar:2003uw, Kawasaki:1991gn, Comelli:2003cv}; a more up to date discussion is presented in Ref.~\cite{Velten:2014nra}) to understand the dark sector coincidence, and an extension was made to the case of neutrinos by Fardon, Nelson, and Weiner \cite{Fardon:2003eh} (called FNW henceforth), followed by Peccei \cite{Peccei:2004sz}. These works predict a time dependent mass for the DM particles and the neutrinos\footnote{Time varying neutrino masses were also discussed in Ref.~\cite{Gu:2003er} in the context of baryogenesis. The authors here considered a higher dimension coupling involving a quintessence field, the Higgs doublet, and neutrinos, as opposed to the Yukawa coupling in FNW.}. In the FNW model a scalar DE field $\varphi$ interacts with the fermionic field of neutrinos through a Yukawa coupling. This model, henceforth referred to as the mass varying neutrino (MaVaN) scenario, is promising to addresses the coincidence problem in the context of neutrinos \cite{Peccei:2004sz} in addition to a straightforward explanation of the neutrino mass beyond the Higgs mechanism of the Standard Model.

A few issues had been pointed out \cite{Afshordi:2005ym,Kaplinghat:2006jk,Brookfield:2006,Brookfield:2008} with the FNW model, one of them being the strong instability of the DE-$\nu$ fluid due to a negative speed of sound, $c_s^2<0$. Various authors have introduced more complexities into the model to get rid of the instabilities. Chitov, August, Natarajan, and Kahniashvili \cite{Chitov:2009ph} (henceforth referred to as CANK) have taken this simple minimal FNW model with the
Ratra-Peebles (referred to as RP henceforth) DE potential \cite{Ratra:1987rm} and analyzed its evolution with the expansion (cooling) of the Universe in the framework of the finite-temperature quantum field theory. This approach allowed these authors \cite{Chitov:2009ph} to systematically derive the effective thermodynamic potential
from the Lagrangian of the model, to eliminate the problem of initial conditions, and to address the problem of stability within the standard analysis of 
(thermodynamically) stable phases.
In this choice of featureless RP potential the neutrino mass arises from an interplay between the bosonic and fermionic contributions, locking the fermionic field around a non-trivial minimum of the thermodynamic potential, regardless of the initial conditions.  Their analysis predicted the existence of stable, metastable, and unstable phases of the combined DE-neutrino fluid. In this approach, the instability mentioned by Afshordi et al. \cite{Afshordi:2005ym} is not an intrinsic flaw of the model. The vanishing velocity of sound is shown to be a physically consistent property of the model at the critical point of the first-order phase transition (known also as spinodal decomposition).  It describes physically the characteristics of the Universe today in a way that is consistent with observations -- the Universe is in an unstable phase below the critical temperature and slowly rolling towards the stable equilibrium (i.e., the true vacuum).

In their work, CANK had performed an analysis for a specific potential, namely the RP potential \cite{Ratra:1987rm}, as mentioned above. In the present work, we extend the analysis to some more potentials, namely the Ferreira-Joyce potential, the inverse exponent potential, and a periodic potential. The rest of the paper is organized as follows: In Sec. \ref{recap} we present a brief overview of the formalism and notations. In Secs. \ref{SecFJ}, \ref{SecIE}, and \ref{SecOsc} we present the analysis for the three different DE potentials. In Sec. \ref{secConcl}, we present our conclusions and discuss the planned future extensions of this work. More technical details on derivations and short overview of CANK \cite{Chitov:2009ph} are given in the Appendices.

All throughout this work, we use the $(+---)$ metric signature, and the natural units $\hbar=c=k_B=1$. To avoid cumbersome notation, a few symbols are reused to denote analogous quantities for separate potentials -- these symbols are to be interpreted in the context of that potential.

\section{Overview of the formalism}
\label{recap}

In this section, we introduce the FNW model  \cite{Fardon:2003eh} for mass varying neutrinos and  review the salient features of the
formalism developed in the paper by CANK  \cite{Chitov:2009ph}. The model consists of a scalar field $\varphi$ which acts as the quintessence dark energy DE; this field has a potential $U(\varphi)$.  The neutrinos are described by the Dirac spinor field $\psi$ with zero chemical potential\footnote{For more details on this point, see Ref.~\cite{Chitov:2009ph}}. The spatially flat Universe is described by the Friedmann-Lema\^{i}tre-Robertson-Walker (FLRW) metric, $ds^2=dt^2-a^2(t)d\mathbf{x}^2$ where $t$ is the cosmic time and $a(t)$ is the scale factor. The latter is related to the total energy density $\rho_\mathrm{tot}$ and total pressure $P_\mathrm{tot}$ through the Friedmann and the continuity equations \cite{Dodelson:2003ft,Weinberg:2008zzc},
\begin{equation}\label{eFried1}
H^2(t)\equiv\left(\frac{\dot{a}}{a}\right)^2=\frac{8\pi G}{3}\rho_\mathrm{tot}~,\quad\quad\quad\dot{\rho}_\mathrm{tot}+\frac{3\dot{a}}{a}\big(\rho_\mathrm{tot}+P_\mathrm{tot}\big)=0~,
\end{equation}
where $H$ is the time-dependent Hubble parameter, and the dot represents a derivative with respect to $t$.

The critical density of the Universe at the present time $t_0$ is $\rho_\mathrm{crit,0}=3H_0^2/8\pi G$
where $H_0$ is the Hubble parameter at the present time, $H_0=H(t)$. It is convenient to express $H_0$ in units of velocity per distance,
$H_0\equiv 100\,h\,\mathrm{km\,s^{-1}\,Mpc^{-1}}$, and the critical density can thus be written as $\rho_\mathrm{crit,0}=8.4\times10^{-11}h^2\,\mathrm{eV^4}$.
It is assumed that the energy density in matter (DM and baryons) accounts for about 30\% of the total energy density, $\Omega_M=\rho_{M,0}/\rho_\mathrm{crit,0}\approx0.3$, and the rest is in the DE-$\nu$ (the contribution from photons can be neglected at the present time, but is important at earlier times), $\Omega_{\varphi\nu,0}=\rho_{\varphi\nu,0}/\rho_\mathrm{crit,0}\approx 0.7$.

The  Lagrangian of the free Dirac field with mass $M_F$ is
\begin{equation}\label{eFermLag}
\mathcal{L}_F=\bar{\psi}\big(i\cancel{\pa}-M_F\big)\psi~,
\end{equation}
where $\cancel{\pa}$ is written in the Feynman slash notation, i.e., $\cancel{\pa}\equiv\gamma^\mu\pa_\mu$, with the $\gamma$'s being the Dirac matrices. Using equilibrium finite-temperature quantum field theory \cite{Kapusta:2006pm}, the pressure $P_F$ and the free energy $F_F$ are found\footnote{Some details are in Sec. II B, C, and D of CANK.} to be
\begin{equation}\label{eFermPress}
P_F=-F_F=-F_0+\frac{1}{3\pi^2}\int_0^\infty\frac{dp\,p^4}{\epsilon(p)}\big[n_F(\epsilon_+)+n_F(\epsilon_-)\big]~,
\end{equation}
where $F_0=-2\int d^3\mathbf{p}\,\epsilon(p)/(2\pi)^3$, $\epsilon(\mathbf{k})=\sqrt{p^2+M_F^2}$, $\epsilon_\pm=\epsilon\pm\mu$, and $n_F(x)=(e^{\beta x}+1)^{-1}$, and $\beta$ is the inverse temperature.

We consider the neutrinos to be initially massless, and they interact with the scalar via a Yukawa coupling of the form,
\begin{equation}\label{eYuk}
\mathcal{L}_\mathrm{coupling}=-g\bar{\psi}\varphi\psi~,
\end{equation}
with $g$ being the coupling constant. We can write the partition function in terms of an effective bosonic action $S^E_\mathrm{eff}(\varphi)$,
\begin{equation}\label{ePartBosEff}
\mathcal{Z}_{\varphi\nu}=\int\mathcal{D}\varphi\,e^{S^E_\mathrm{eff}(\varphi)}=\int\mathcal{D}\varphi\,e^{S^E_B+\log\det \hat{D}(\varphi)}~,
\end{equation}
where,
\begin{equation}\label{eDirOp}
\hat{D}(\varphi)=-\beta\left[\frac{\pa}{\pa \tau}-i\frac{\gamma^0\bm{\gamma}\cdot\nabla}{a}+\gamma^0g\varphi-\mu\right],
\end{equation}
and $S^E_B$ is the Euclidean action of the bosonic field $\varphi$. If the effective action $S^E_\mathrm{eff}(\varphi)$ is minimized at the constant value $\varphi=\varphi_m$, which is the expectation value of the field $\varphi_m=\langle\varphi\rangle$, then from Eq. \eqref{eDirOp} we see that the fermion gets an effective mass,
\begin{equation}\label{eFermMass}
m=g\varphi_m~.
\end{equation}
At $\varphi=\varphi_m$, we can write,
\begin{equation}\label{ePartSad}
\mathcal{Z}_{\varphi\nu}=\mathcal{Z}_F\,e^{-\beta V U(\varphi_m)}~,
\end{equation}
where $\mathcal{Z}_F$ is the partition function for massless fermions ($M_F=0$). The free energy $F_{\varphi\nu}$ of the DE-neutrino fluid is then given by $F_{\varphi\nu}(\varphi_m)=U(\varphi_m)+F_F(\varphi_m)$. This is a saddle-point approximation, and it is self-consistent if the thermodynamic potential is minimized at $\varphi=\varphi_m$,
\begin{equation}\label{eThermSaddle}
\left.\frac{\pa F_{\varphi\nu}(\varphi)}{\pa\varphi}\right|_{\mu,\beta;\varphi=\varphi_m}=0~,\quad\quad\quad \left.\frac{\pa^2F_{\varphi\nu}(\varphi)}{\pa\varphi^2}\right|_{\mu,\beta;\varphi=\varphi_m}>0~.
\end{equation}
Taking the derivative of Eq. \eqref{ePartSad} at $\varphi_m$, and using the first of Eq. \eqref{eThermSaddle}, we get the \textit{fermionic mass  equation},
\begin{equation}\label{eGap}
U'(\varphi_m)+g\rho_s=0~,
\end{equation}
where $\rho_s$ is the \textit{scalar fermionic density} (or the \textit{chiral condensate density}),
\begin{equation}
\label{eRhoS}
\rho_s=\frac{m}{\pi^2}\int_0^\infty\frac{dp\,p^2}{\epsilon(p)}\big[n_F(\epsilon_+)+n_F(\epsilon_-)-1\big]~.
\end{equation}
The potentials are redefined with respect to  their vacuum values, i.e., $F_F\mapsto F_F-F_0$ (which is equivalent to redefining $P_F\mapsto P_F-P_0$ with $P_0=-F_0$) and $\rho_s\mapsto \rho_s-\rho_0$ with $\rho_0=-(m/\pi^2)\int_0^\infty dp\,p^2/\epsilon(p)$. One can rewrite the integrals in Eqs. \eqref{eFermPress} and \eqref{eRhoS} in a more amenable form by the substitution $z=\beta\epsilon$. In what follows we adopt the notations for dimensionless parameters used in CANK,
\begin{equation}
\label{eDimLess}
\kappa \equiv \frac{g\varphi}{T}~,\quad\quad\quad
\Delta\equiv\frac{M}{T}~,\quad\quad\quad
\quad F_R\equiv\frac{F_{\varphi\nu}}{M^4}~,
\end{equation}
where $M$ is the mass scale of the potential, to be specified below for each case considered.
In addition we set the dimensionless Yukawa coupling g = 1 in the following, which is equivalent to some rescaling of the various dimensionful parameters\cite{Chitov:2009ph}.
Then we obtain
\begin{equation}\label{eFermPres2RhoS2}
P_F=\frac{2N_F}{3\pi^2\beta^4}\mathscr{I}_{3/2}(\kappa), \quad\quad\quad
\rho_s=\frac{2N_Fm}{\pi^2\beta^2}\mathscr{I}_{1/2}(\kappa)~,
\end{equation}
where
\begin{equation}\label{eI1I2}
\mathscr{I}_\nu (\kappa) \equiv \int_{\kappa}^\infty \frac{dz\,\big(z^2-\kappa^2 \big)^\nu }{e^z+1}~.
\end{equation}
For Eq. \eqref{eGap} to have a solution, the potential $U(\varphi)$ should be decreasing, at least in some range of parameters.  For all DE potentials satisfying the above condition, there are only certain ranges of parameters (phases) for which a viable solution for
nontrivial minimum of $F_{\varphi\nu}$ exists. The interested reader is referred to the end of Sec. III of CANK for a more detailed discussion. Some results for the RP potential are provided in Appendix \ref{eAppRP} for comparison with our results for other potentials, and also as a useful refresher. For all three quintessence potentials considered in this paper, as well as the RP potential, we find that the interaction between the fermionic and scalar fields results in new nontrivial minima of the coupled thermodynamic potentials.

Along with the fermionic mass defined above, we will need the mass of the scalar field.
$F_{\varphi\nu}(\varphi)$ is the finite-temperature counterpart of the effective potential known from the quantum field theory
at $T=0$ (see Ref.~\cite{Ryder:1985wq}, for example). The mass of the field $\varphi$ is defined as
\begin{eqnarray}
\label{eRenormMass}
 m_{\varphi}^2 \equiv \left.\frac{\pa^2 F_{\varphi\nu}(\varphi)}{\pa\varphi^2}\right|_{\varphi=\varphi_m}
=\left(\frac{\pa^2 U(\varphi)}{\pa\varphi^2}-\frac{2N_F}{3\pi^2\beta^2}\mathscr{I}_{3/2}''(\kappa)\right) \Bigg|_{\varphi=\varphi_m}~.
\end{eqnarray}
We note that this contains a fermionic contribution which ``renormalizes" the ``bare" value due to the scalar potential contribution $U''(\varphi)$.

The adiabatic sound velocity $c_s$ for the $\varphi\nu$ fluid is found from the definition
\begin{equation}
\label{eSoundSpeed1}
c_s^2\equiv\frac{dP_{\varphi\nu}}{d\rho_{\varphi\nu}}~,
\end{equation}
where $P_{\varphi\nu}=-F_{\varphi\nu}$ is the pressure and $\rho_{\varphi\nu}=U(\varphi)+2N_F\int_0^\infty dp\,p^2\,n_F(\epsilon)\,\epsilon/\pi^2$ is the energy density of this fluid. In terms of $\kappa$ we can write
\begin{equation}
\label{ePhiNuEnergy}
\rho_{\varphi\nu}=U(\varphi)+\frac{2N_F}{\pi^2\beta^4} \mathscr{I}_\varepsilon (\kappa)~,
\end{equation}
where
\begin{equation}
\label{Ieps}
  \mathscr{I}_\varepsilon (\kappa) \equiv
   \int_{\kappa}^\infty \frac{dz\, z^2\big(z^2-\kappa^2 \big)^{1/2} }{e^z+1}~.
\end{equation}
The definition in Eq. \eqref{eSoundSpeed1} yields \cite{Chitov:2009ph}
\begin{equation}
\label{cs2Eq}
    c_s^2 = \frac{\partial P/\partial \Delta}{(\partial \rho/\partial \Delta)+
    (\partial \rho/\partial \kappa)\dot{\kappa}_m }\Bigg\vert_{\kappa=\kappa_m}~,
\end{equation}
where
\begin{equation}
\label{Dkap}
\dot{\kappa}_m \equiv \frac{d \kappa}{ d \Delta} \Big\vert_{\kappa=\kappa_m},
\end{equation}
and $\kappa$ is related to $\Delta$ through the mass equation, Eq. \eqref{eGap}.
For an arbitrary quintessence potential which can be written as a function of $m/M \equiv \kappa /\Delta$ and in the range of parameters where a non-trivial minimum  satisfying Eq. \eqref{eGap} exists, the equation for the speed of sound, Eq. \eqref{cs2Eq}, can be written in the form most convenient for calculations as
\begin{equation}
\label{eCsSqInteg}
c_s^2=\left.\frac{4\mathscr{I}_{3/2}(\kappa)/3 + \kappa^2\mathscr{I}_{1/2}(\kappa)}{
4\mathscr{I}_\varepsilon(\kappa)-\kappa^2\mathscr{I}_{1/2}(\kappa)+
\left[\kappa\mathscr{I}_{1/2}(\kappa)-\mathscr{I}_\varepsilon'(\kappa)\right] \Delta \dot{\kappa}_m}\right|_{\kappa=\kappa_m}~.
\end{equation}
In the above, $\dot{\kappa}_m$, determined from the minimum equation Eq. \eqref{eGap}, is the only \textit{explicitly} potential dependent term in the above expression.

The equation of state for the $\varphi\nu$ fluid is parameterized via
\begin{equation}\label{eEoS1}
w_{\varphi\nu}\equiv\frac{P_{\varphi\nu}}{\rho_{\varphi\nu}}~.
\end{equation}
In the following we will need high- and low-temperature asymptotes of the above fermionic integrals. One can find
\begin{equation}
\label{I32As}
     \mathscr{I}_{3/2}(\kappa)=  \left\{
                \begin{array}{ll}
    \frac{7 \pi^4}{120}- \frac{\pi^2}{8} \kappa^2+\mathcal{O}(\kappa^4\log \kappa)~, & \kappa < 1 \\[0.3cm]
    3 \kappa^2 K_2(\kappa)+\mathcal{O}(e^{-2 \kappa}) ~, & \kappa \gtrsim 1
                \end{array},
       \right.
\end{equation}
where $K_\nu (x)$ is the modified Bessel function of the second kind. Similarly,
\begin{equation}
\label{I12As}
     \mathscr{I}_{1/2}(\kappa)=  \left\{
                \begin{array}{ll}
    \frac{\pi^2}{12}- \frac14 \kappa^2 \log \kappa -(\frac18 + \frac{\log \pi -\gamma}{4}) \kappa^2 +\\[0.3cm]
    +  \mathcal{O}(\kappa^3\log \kappa)~, & \kappa < 1 \\[0.3cm]
    \kappa  K_1(\kappa)+\mathcal{O}(e^{-2 \kappa}) ~, & \kappa \gtrsim 1
                \end{array},
       \right.
\end{equation}
where $\gamma$ is Euler's gamma constant. We will also need
\begin{equation}
\label{IepsAs}
    \mathscr{I}_\varepsilon(\kappa) =  \left\{
                \begin{array}{ll}
    \frac{7 \pi^4}{120}- \frac{\pi^2}{24} \kappa^2+\mathcal{O}(\kappa^4 \log \kappa)~, & \kappa < 1 \\[0.3cm]
    3 \kappa^2 K_2(\kappa)+\kappa^3 K_1(\kappa)+ \mathcal{O}(e^{-2 \kappa}) ~, & \kappa \gtrsim 1
                \end{array}.
       \right.
\end{equation}

\section{The Ferreira-Joyce potential}
\label{SecFJ}

This potential of the form
\begin{equation}
\label{eFJPot}
U(\varphi) = M^4 e^{-\varphi/M_{\varphi}}
\end{equation}
was proposed in Ref.~\citep{Ferreira:1997hj} and discussed in Refs.~\cite{Brookfield:2006,Amendola:2015ksp}.
In this work we do not preset the scales for $M$ and $M_\varphi$, and treat them as two independent adjustable mass parameters required to consistently yield the neutrino masses and the observed value of the present DE density\footnote{The scale $M_\varphi$ is typically taken to be the Planck mass, see for example Ref.~\cite{Avsajanishvili:2017zoj}. In this work the fitting yields $M_\varphi \gtrsim \mathcal{O}(\mathrm{eV})$.}.
In addition to the ones in Eq. \eqref{eDimLess}, we define a single dimensionless parameter $\lambda$,
\begin{equation}
\label{lam}
\lambda\equiv M/M_\varphi~,
\end{equation}
which incorporates two mass scales. This parameter controls the validity of our results, and the approximations used.
We will thus be dealing with the dimensionless thermodynamic potential
\begin{equation}\label{eFRFJ}
F_R=e^{-\kappa\lambda/\Delta}-\frac{2N_F}{3\pi^2 \Delta^4} \mathscr{I}_{3/2}(\kappa)~.
\end{equation}
The minimum (mass) equation, Eq. \eqref{eGap}, reads
\begin{equation}
\label{eGapFJ}
\frac{\pi^2\lambda}{2N_F}\Delta^3=\kappa\,e^{\kappa\lambda/\Delta}\mathscr{I}_{1/2}(\kappa)
\equiv\mathcal{I}_\Delta~.
\end{equation}
We denote as $\kappa_m$ the value of $\kappa$ which solves \eqref{eGapFJ} and minimizes $F_R$.

\begin{figure}[h!]
\begin{center}
\includegraphics[width=0.6\textwidth]{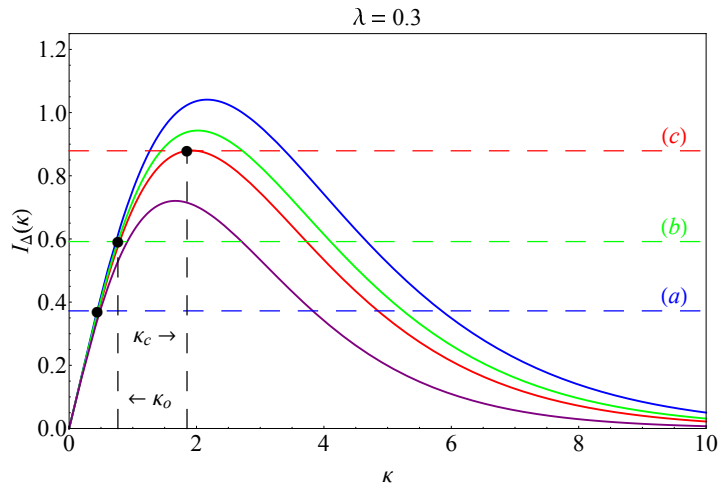}
\end{center}
\caption[]{Thermodynamically stable solutions ($\kappa_m$) of Eq. \eqref{eGapFJ} for $\lambda=0.3$ and different values of $\Delta$, showing a stable (blue) state, the metastable (green) state, the critical point (red), and a state after the phase transition (purple). The dashed line corresponding to the left hand side of Eq. \eqref{eGapFJ} for the purple $\mathcal{I}_\Delta$ curve lies beyond the range of the plot. Here $\kappa_\mathrm{c}$ is the solution of the gap equation at $\Delta=\Delta_\mathrm{c}$, and $\kappa_o$ is that at $\Delta=\Delta_o$.}
\label{FigIDelFJ}
\end{figure}

\begin{figure}[h]
\begin{center}
\includegraphics[width=0.6\textwidth]{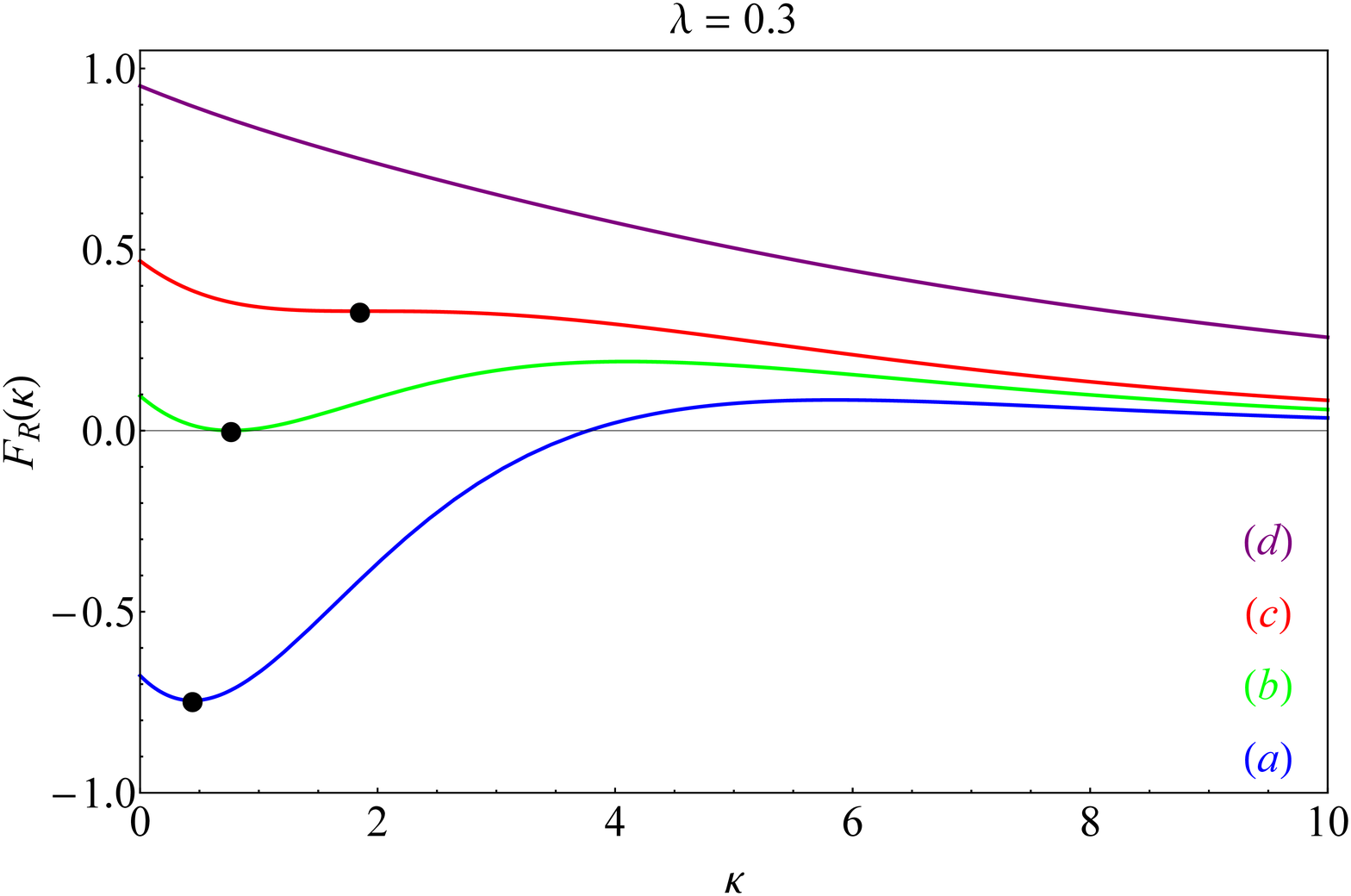}
\end{center}
\caption{The thermodynamic potential, $F_R$, as given by Eq. \eqref{eFRFJ} is plotted for the cases in Fig. \ref{FigIDelFJ}. The stable solutions of Eq. \eqref{eGapFJ} minimize the corresponding potentials, and are indicated here by big filled dots.}
\label{FigFRFJ}
\end{figure}

As opposed to the case of the RP potential, the right hand side of the mass equation is $\Delta$-dependent (compare with Eq. \eqref{eGapRP}). In Fig.~\ref{FigIDelFJ}, we plot the integral $\mathcal{I}_\Delta(\kappa)$ by choosing a fiducial value of $\lambda=0.3$ and various values of $\Delta$, and in Fig.~\ref{FigFRFJ}, we plot the thermodynamic potential in Eq. \eqref{eFRFJ} for the corresponding cases. The red curves denote the critical temperature (we call it $\Delta=\Delta_\mathrm{c}$) when the local minimum of the potential $F_R$ disappears (and Eq. \eqref{eGapFJ} no longer has a nontrivial solution). The blue curves denote the stable phase $\Delta<\Delta_\mathrm{c}$, and the purple curve is after the phase transition, $\Delta>\Delta_\mathrm{c}$, when the minimum equation has no solution. For $\Delta>\Delta_\mathrm{c}$, i.e., in the unstable phase, the field $\varphi$ rolls towards the global minimum at $\varphi\rightarrow\infty$. The green curves denote the point of metastability, $\Delta=\Delta_0$, when the pressure of the $\varphi\nu$ fluid vanishes (see Fig.~\ref{FigFRFJ}). The dashed horizontal lines in Fig.~\ref{FigIDelFJ} denote the right hand side of Eq. \eqref{eGapFJ} for the corresponding values of $\Delta$.

\begin{figure}[h]
\begin{center}
\includegraphics[width=0.6\textwidth]{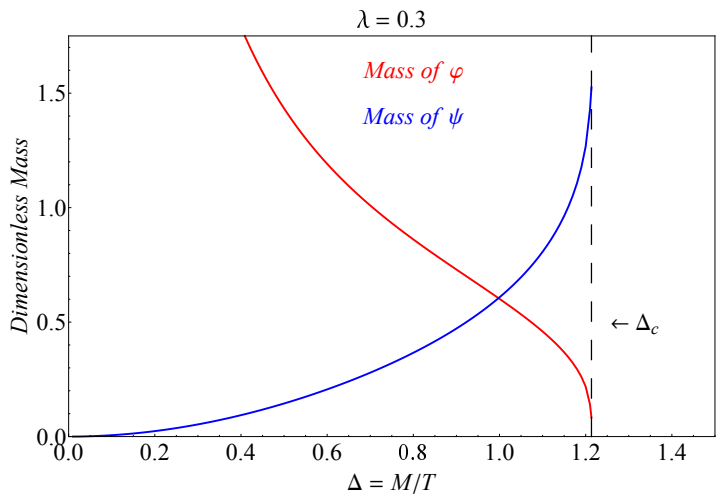}
\end{center}
\caption{Temperature dependence of the fermion and scalar masses for the Ferreira-Joyce potential with $\lambda=0.3$}
\label{FigMassesFJ}
\end{figure}

To estimate the critical point, we first need to determine the local maximum of $\mathcal{I}_\Delta$; the value of the integral $\mathcal{I}_\Delta$ at this point should satisfy the mass equation Eq. \eqref{eGapFJ}. Making the low-temperature approximation to $\mathcal{I}_\Delta$, we evaluate the critical quantities,
\begin{equation}\label{eCritPtFJ}
\kappa_\mathrm{c}\approx\frac{3}{2}\dfrac{\sqrt{1+B/\lambda^{8/3}}+1}{\sqrt{1+B/\lambda^{8/3}}-1},\quad\quad\quad
\Delta_\mathrm{c}=\frac{\lambda}{2}\left(1+\sqrt{1+B/\lambda^{8/3}}\right)~,
\end{equation}
where $B=6\left(2N_F^2\right)^{1/3}/(\pi e)$. Note that these are rough estimates, and for all our plots we use the exact numerical solutions. As for the RP potential, we can derive an approximate expression for the fermion mass for the stable phase $\Delta<\Delta_\mathrm{c}$. At very high temperatures $T \gg M$ the approximation $\mathcal{I}_\Delta(\kappa)\simeq\pi^2\kappa/12$ yields the fermion mass
\begin{equation}
\label{eFermMassFJ}
m \approx \frac{6\lambda}{N_F} \frac{M^3}{T^2} ~,
\end{equation}
and the mass of the scalar field in this regime is
\begin{equation}
\label{eRenMassFJ}
m_\varphi \approx \frac{N_F}{6} \frac{T^2}{M}~.
\end{equation}
It is easily checked that just as for the case of the RP potential (see Eq. (65) of CANK), the fermion mass approaches the critical value as $m_\mathrm{c}-m\sim (T-T_\mathrm{c})^{1/2}$. Indeed near the critical point,
\begin{equation}
\label{eCritKappaDeltaFJ}
1-\left(\frac{T_\mathrm{c}}{T}\right)^3=\frac{3}{4}\left(\frac{\kappa_m}{\kappa_\mathrm{c}}-1\right)^2~,
\end{equation}
from which the above critical exponent is calculated in a straightforward manner. The scalar field mass vanishes at the critical point, since
this is an inflection point of the thermodynamic potential.
The masses $m$ and $m_\varphi$ in units of $M$ are plotted in Fig. \ref{FigMassesFJ}.
To analyze the sound velocity, we derive the explicit formula for
\begin{equation}
\label{DLogDKFJ}
   \frac{d \log \kappa_m}{d \log \Delta } = \frac{3 +\lambda \kappa/ \Delta}{1+ \lambda \kappa/ \Delta +d
    \log \mathscr{I}_{1/2}(\kappa)/d \log \kappa} \Bigg\vert_{\kappa=\kappa_m}
\end{equation}
from Eq. \eqref{eGapFJ} and use it in Eq. \eqref{eCsSqInteg}.
We find that everywhere at $T> T_{\mathrm{c}}$, including the stable and metastable massive
phases  $c_s^2>0$ ($w_{\varphi\nu}<0$ for the latter) the model is stable with respect to the density fluctuations.

At high temperatures $\Delta \ll 1$ and $\kappa_m \ll 1$ one can find from  Eq. \eqref{eCsSqInteg}, using the asymptotes
in Eqs. \eqref{I32As}, \eqref{I12As}, and \eqref{IepsAs}, that $c_s^2 \approx 1/3$,
as expected for a relativistic gas. The sound velocity decreases with decreasing temperature, and it vanishes at $T \to T_{\mathrm{c}}^+$.
Qualitatively, the vanishing speed of sound is due to divergent $\dot{\kappa}_m$ at the critical point.
We have seen that near the critical point $m-m_\mathrm{c} \propto (T-T_\mathrm{c})^{1/2}$, thus
\begin{equation}\label{emDerNearCrit}
\left.\dfrac{d \kappa_m}{d \Delta}\right|_{\kappa=\kappa_\mathrm{c}} \propto \left.\frac{\pa m}{\pa T}\right|_{T_\mathrm{c}} \propto \frac{1}{\sqrt{T-T_\mathrm{c}}}\rightarrow\infty~.
\end{equation}
The sound velocity, therefore, drops to 0 at the critical point.

\begin{figure}[h]
\begin{center}
\includegraphics[width=0.6\textwidth]{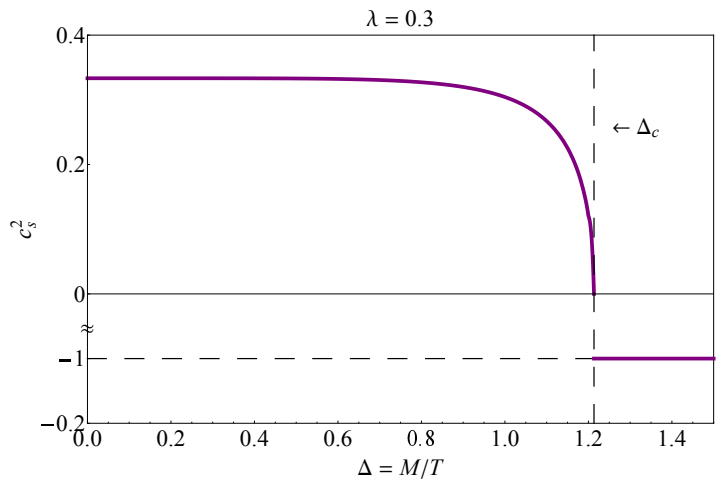}
\end{center}
\caption{Temperature dependence of the speed of sound $c_s^2$ in the $\varphi\nu$ fluid for the Ferreira-Joyce potential with $\lambda=0.3$.}
\label{FigCs2FJ}
\end{figure}

After the phase transition, there is no nontrivial solution of Eq. \eqref{eGapFJ} for the minimum, and $\dot{\kappa}_m$ in the defining
Eq. \eqref{cs2Eq} is absent. The stable solution is the global minimum of $F_R$ at $\kappa\rightarrow \infty$, and from
Eq. \eqref{cs2Eq} we formally obtain $c_s^2=-1$ at $T<T_c$. The exact numerical calculation of $c_s^2$ for temperatures $T>T_c$ is plotted in Fig. \ref{FigCs2FJ}.
Note however that the derivation within the equilibrium approach leading to the exact value $c_s^2=-1$  may be questioned for the temperatures $T<T_c$.
However the imaginary sound velocity below $T_c$ indicates correctly the physical effect of growing density fluctuations due to instability (also known as clustering according to Wetterich and collaborators \cite{Baldi:2011es, Ayaita:2011ay, Ayaita:2014una, Casas:2016duf})
which are expected to occur in a system going through the first order phase transition.

\subsection{Dynamics after the transition}
\label{secFJDynAft}
After the phase transition the evolution of the coupled $\varphi \nu$ model cannot be described by small oscillations around the stable solutions obtained by minimization, and its dynamics is governed by the full equation of motion, Eq. \eqref{eEOMRP}, supplemented with the Friedmann equations.
It was shown by CANK from numerical analyses that the exact numerical solution of the dynamics governed by Eq. \eqref{eEOMRP} oscillates around the mean solution given by setting the right hand side of Eq. \eqref{eEOMRP} to zero.
For the Ferreira-Joyce potential we get, following this analytic method,
\begin{equation}
\label{eFJRhoSCrit}
\frac{\rho_{s,\mathrm{c}}}{M^3}=\lambda\exp\left(-\frac{3}{R-1}\right)~,
\end{equation}
where $R=\sqrt{1+B/\lambda^{8/3}}$. The mean $\varphi$ after the transition, denoted by $\bar{\varphi}$, is given by
\begin{equation}
\label{eFJRhoAfter}
e^{-\lambda\bar{\varphi}/M}=e^{-3/(R-1)}\left(\frac{T}{T_\mathrm{c}}\right)^3~,
\end{equation}
where $M^4 e^{-3/(R-1)}$ is the energy density in the $\varphi\nu$ fluid at the phase transition. As in CANK, we set a value of one of the two independent parameters of the model, i.e., $M$, relating it to DE density at the present time $\rho_{\varphi,\mathrm{now}}$ getting thus
\begin{equation}
\label{eMValFJ}
M(\lambda) =\frac{\rho_{\varphi,\mathrm{now}}}{(T_\mathrm{now}\Delta_\mathrm{c})^3}\,e^{3/(R-1)}=
\frac{8\rho_{\varphi,\mathrm{now}}}{\big(T_\mathrm{now}\lambda(1+R)\big)^3}\,e^{3/(R-1)}~,
\end{equation}
where $T_\mathrm{now}$ is the present temperature. We evaluate the current neutrino mass from
\begin{equation}\label{emNowFJ}
m_\mathrm{now}(\lambda) =\varphi_\mathrm{now}=\frac{M}{\lambda}\log \frac{M^4}{\rho_{\varphi,\mathrm{now}}}~.
\end{equation}
In Tab. \ref{TabFJMass}, we list the values of $M$ and $m_\mathrm{now}$ for various choices of $\lambda$.
\begin{table}[h]
\centering
\begin{tabular}{l|l|l}
\hline
$\bm{\lambda}$ & $\bm{M\,\mathrm{(eV)}}$  & $\bm{m_\mathrm{now}\,\mathrm{(eV)}}$ \\
\hline
$1$ & $74.37$  & $3085.25$ \\
$10^{-1}$ & $0.73$  & $167.28$ \\
$10^{-2}$ & $0.07$  & $99.77$ \\
$10^{-3}$ & $0.007$  & $32.76$ \\
$4\times10^{-4}$ & $0.0029$  & $6.38$ \\
$3.25\times10^{-4}$ & $0.0024$  & $0.35$ \\
\hline
\end{tabular}
\caption{Approximate neutrino masses at the present epoch for various choices of $\lambda$ for the Ferreira-Joyce potential.}
\label{TabFJMass}
\end{table}
One can see that the reasonable neutrino masses consistent with the current DE density are obtained in the range of the model parameters
$M \sim 0.2 \times 10^{-3}$ eV and $M_\varphi =M /\lambda  \sim 6-7\,\mathrm{eV}$.

\begin{figure}[h]
\begin{center}
\includegraphics[width=0.6\textwidth]{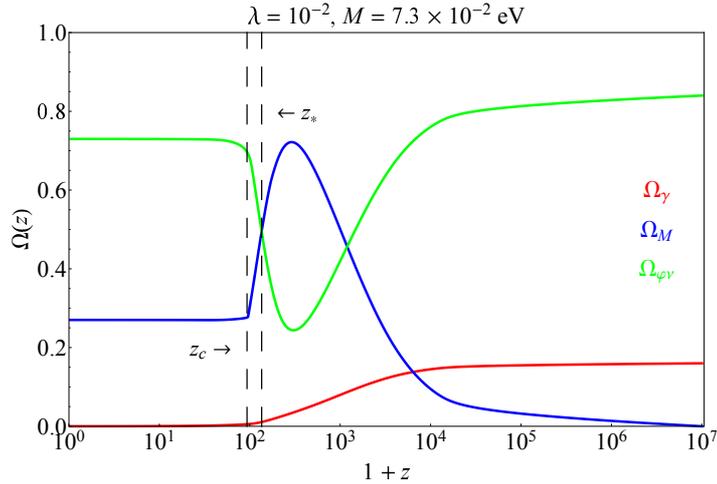}
\caption{The relative densities of photons, matter, and the $\varphi\nu$ fluid as a function of redshift $z$ for the case of the Ferreira-Joyce potential. Here, $z_\star\approx 134.38$ denotes the epoch of matter-DE equality, and $z_\mathrm{c}\approx 92.22$ denotes the redshift at which the temperature crosses $T_\mathrm{c}$. Note that we are plotting the temperatures for $1+z\leq 10^7$, i.e., $T\lesssim 2.35\,\mathrm{KeV}$, that is, \textit{well below} the epoch of $e^--e^+$ annihilation which corresponding to $T\sim 0.5\,\mathrm{MeV}$. As discussed above, the late time behavior is in contradiction to the observed evolution of the Universe.}
\label{FigOmeFJ}
\end{center}
\end{figure}

\begin{figure}[h]
\begin{center}
\includegraphics[width=0.6\textwidth]{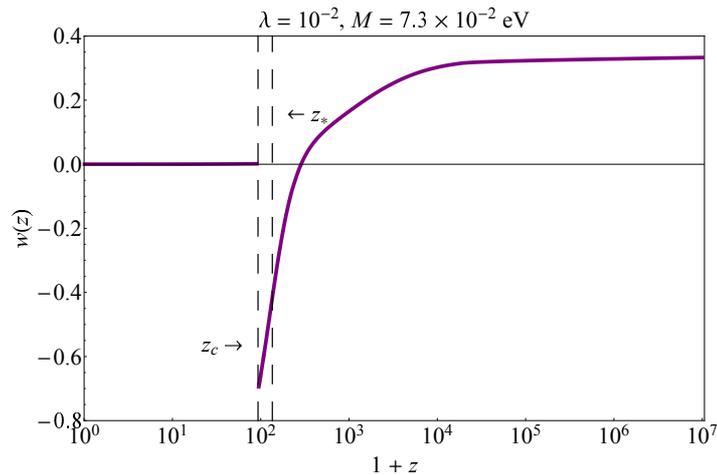}
\caption{Equation of state for the entire Universe as a function of redshift for the Ferreira-Joyce potential. The redshift range is the same as in Fig. \ref{FigOmeFJ}. It is clearly seen that the net equation of state at late times is unable to produce an exponential expansion.}
\label{FigEosZFJ}
\end{center}
\end{figure}

In Fig. \ref{FigOmeFJ}, we show the relative energy densities of the various components of the Universe, and in Fig. \ref{FigEosZFJ}, we show the evolution of the equation of state parameter of the entire Universe. Note that as opposed to the choice of $\lambda=0.3$ for the earlier diagrams in this section, we have now chosen $\lambda=0.01$ (giving us $M=0.073\,\mathrm{eV}$) purely for visual purposes. For completeness, we mention here that for a component $I$, the relative energy density is defined as $\Omega_I(z)\equiv \rho_I(z)/\rho_\mathrm{tot}(z)$ with $z$ being the cosmological redshift and $\rho_\mathrm{tot}(z)$ is the total energy; similarly the equation of state parameter of the entire Universe is $w_\mathrm{tot}\equiv P_\mathrm{tot}(z)/\rho_\mathrm{tot}(z)$.

We note from Fig. \ref{FigOmeFJ}, as well as from Eq. \eqref{eFJRhoAfter} that the $\varphi\nu$ fluid essentially behaves like a nonrelativistic matter component after the phase transition. Even though it leads to viable masses for neutrinos in a certain range of the parameter space, the Ferreira-Joyce potential
is unable to produce an exponential expansion at late times.  This rules out the use of this potential for the MaVaN scenario.

\section{The inverse exponent potential}
\label{SecIE}

\begin{figure}[h]
\begin{center}
\includegraphics[width=0.6\textwidth]{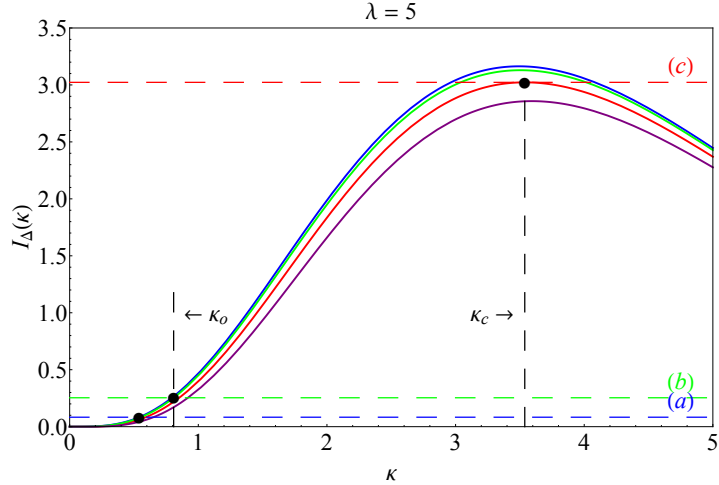}
\caption{Thermodynamically stable solutions ($\kappa_m$) of Eq. \eqref{eGapIE} for $\lambda=5$ and different values of $\Delta$, showing a stable (blue) state, the metastable (green) state, the critical point (red), and a state after the phase transition (purple) for which the dashed line denoting the LHS of Eq. \eqref{eGapIE} lies beyond the range of the plot. Here again $\kappa_\mathrm{c}$ is the solution of the gap equation at $\Delta=\Delta_\mathrm{c}$, and $\kappa_o$ is that at $\Delta=\Delta_o$.}
\label{FigIEGap}
\end{center}
\end{figure}

\begin{figure}[h]
\begin{center}
\includegraphics[width=0.6\textwidth]{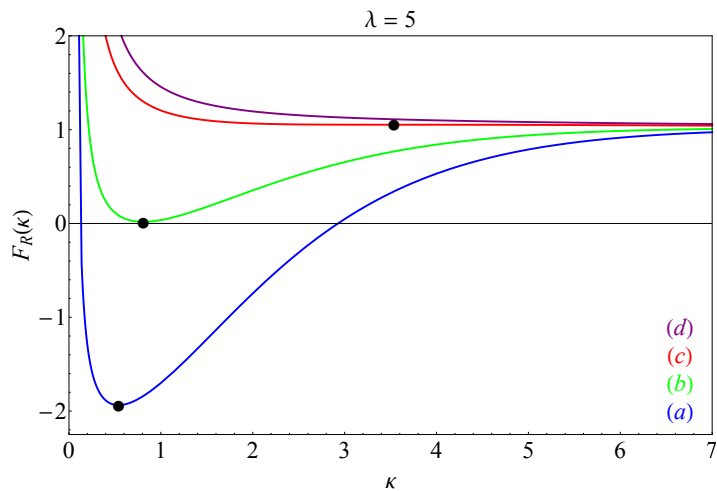}
\caption{The thermodynamic potential, $F_R$, as given by Eq. \eqref{eThermPotIE} is plotted for the cases in Fig. \ref{FigIEGap}. The stable solutions of Eq. \eqref{eGapIE} minimize the corresponding potentials, and are indicated here by big filled dots.}
\label{FigIEFR}
\end{center}
\end{figure}

Potentials like these were discussed by Caldwell and Linder \citep{Caldwell:2005tm}. We take it in the form,
\begin{equation}\label{eIEPot}
U(\varphi)=M^4\,e^{M_\varphi/\varphi}~,
\end{equation}
where $M$ and $M_\varphi$ are two independent mass parameters, analogous to those for the Ferreira-Joyce potential.
Using the dimensionless notations in Eqs. \eqref{eDimLess} and \eqref{lam}, the normalized thermodynamic potential is,
\begin{equation}\label{eThermPotIE}
F_R=e^{\Delta/\kappa\lambda}-\frac{2N_F}{3\pi^2 \Delta^4} \mathscr{I}_{3/2}(\kappa)~,
\end{equation}
and the fermion mass equation becomes\footnote{Note that the $\mathcal{I}_{\Delta}$ defined above in Eq. \eqref{eGapIE} is not the same as the integral with the same name used in Sec. \ref{SecFJ}; we merely reuse the notations.},
\begin{equation}
\label{eGapIE}
\frac{\pi^2}{2N_F\lambda}\Delta^5=e^{-\Delta/\kappa\lambda}\kappa^3\mathscr{I}_{1/2}(\kappa)\equiv\mathcal{I}_\Delta(\kappa)~.
\end{equation}
We plot the integral $\mathcal{I}_\Delta(\kappa)$ for a fiducial value of $\lambda=5$ and various values of $\Delta$ in Figs. \ref{FigIEGap}, and in Fig. \ref{FigIEFR}, we plot the thermodynamic potential in Eq. \eqref{eThermPotIE} for the corresponding cases.

\begin{figure}[h]
\begin{center}
\includegraphics[width=0.6\textwidth]{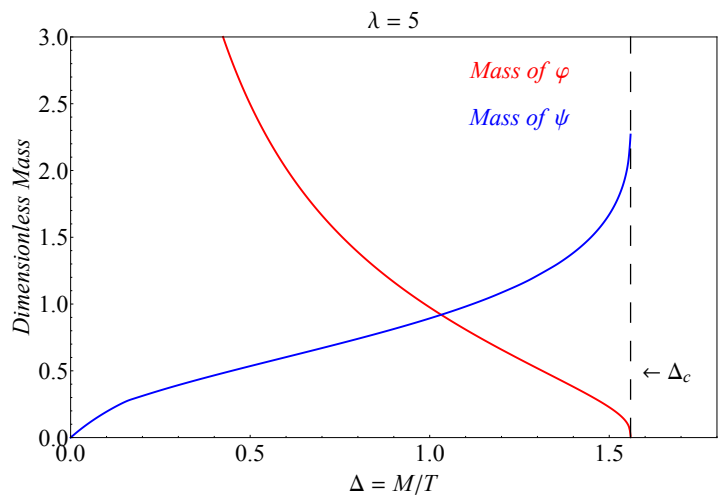}
\caption{Temperature dependence of the fermion and scalar masses for the inverse exponent potential with $\lambda=5$.}
\label{FigMassesIE}
\end{center}
\end{figure}

Since the algebraic analysis of the minimum equation Eq. \eqref{eGapIE} and related results for this potential are quite involved, we have relegated this material to the Appendix \ref{eAppIE}. The outcome for the stable and metastable phases for this model are qualitatively similar to the case of the Ferreira-Joyce potential above. Analogous to Eq. \eqref{eCritKappaDeltaFJ}, we have,
\begin{equation}\label{eCritKappaDeltaIE}
1-\left(\frac{T_\mathrm{c}}{T}\right)^5=\left(\frac{7}{4}+\frac{\Delta_\mathrm{c}}{\lambda\kappa_\mathrm{c}}\right)\left(\frac{\kappa_m}{\kappa_\mathrm{c}}
-1\right)^2~,
\end{equation}
where the approximate analytic expressions for the critical values on the right hand side are given in Eqs. \eqref{e205} and \eqref{e206} respectively. As before, the fermion mass approaches the critical value as $m_\mathrm{c}-m\sim (T-T_\mathrm{c})^{1/2}$, while the scalar field mass vanishes at the critical temperature $T_c$. The exact numerical solutions of the equations for the fermion and the scalar field masses are plotted in Fig. \ref{FigMassesIE}.

\begin{figure}[h]
\begin{center}
\includegraphics[width=0.6\textwidth]{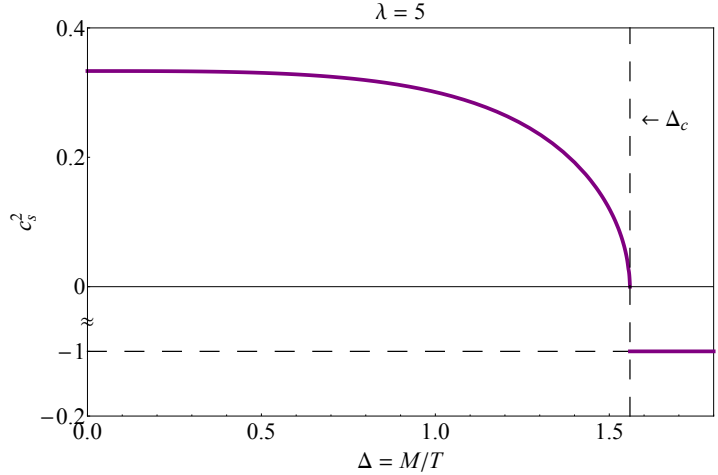}
\caption{Temperature dependence of the speed of sound $c_s^2$ in the $\varphi\nu$ fluid for the inverse exponent potential with $\lambda=5$.}
\label{FigCs2IE}
\end{center}
\end{figure}

To calculate the speed of sound we found
\begin{equation}
\label{DLogDKIE}
   \frac{d \log \kappa_m}{d \log \Delta } = \frac{5 + \Delta/\lambda \kappa}{1+ \Delta/\lambda \kappa
    +d \log \mathscr{I}_{1/2}(\kappa)/d \log \kappa} \Bigg\vert_{\kappa=\kappa_m}
\end{equation}
from Eq. \eqref{eGapIE} and used it in Eq. \eqref{eCsSqInteg}.
The model is stable with respect to the density fluctuations everywhere at $T> T_{\mathrm{c}}$ (including the metastable
phase where $w_{\varphi \nu}<0$), since  $c_s^2>0$. Similarly to the case of the Ferreira-Joyce potential, the parameter
$\dot{\kappa}_m \to \infty$ is divergent at the critical point $T_c$ of the first-order phase transition, so $c_s^2 \to 0$ as $T \to T_c^+$. As explained
for the Ferreira-Joyce potential, the equilibrium thermodynamics yields $c_s^2=-1$ for $T<T_c$ after the transition.
The numerical results for the speed of sound are shown in Fig. \ref{FigCs2IE}.

\subsection{Dynamics after the transition}

Similarly to Eq. \eqref{eFJRhoSCrit} we derive the following expression for the critical scalar density,
\begin{equation}\label{eIERhoSCrit}
\frac{\rho_{s,\mathrm{c}}}{M^3}=\frac{\lambda}{S^2}e^{1/S}~,
\end{equation}
where $S=2/7+7e^{7/5}(a\lambda^4)^{1/5}/2$; the critical energy density is given by,
\begin{equation}\label{eIERhoCrit}
\frac{\rho_{\varphi,\mathrm{c}}}{M^4}=e^{1/S}~.
\end{equation}
The mean solution for $\varphi$ (about which the exact solution is expected to rapidly fluctuate, as mentioned previously) is given by,
\begin{equation}\label{eIERhoAfter}
\left(\frac{M}{\lambda\bar{\varphi}}\right)^2\,e^{M/\lambda\bar{\varphi}}=\frac{e^{1/S}}{S^2}\left(\frac{T}{T_\mathrm{c}}\right)^3~.
\end{equation}
As in the previous section, we make an estimate for $M$,
\begin{equation}\label{eMValIE}
M(\lambda)=\frac{S^2e^{-1/S}\rho_{\varphi,\mathrm{now}}}{(T_\mathrm{now}\Delta_\mathrm{c})^3}\log^2\frac{\rho_{\varphi,\mathrm{now}}}{M^4}~.
\end{equation}
and the neutrino mass at the present time,
\begin{equation}\label{emNowIE}
m_\mathrm{now}(\lambda)=\bar{\varphi}_\mathrm{now}=\frac{M}{\lambda}\displaystyle{\frac{1}{\log(\rho_{\varphi,\mathrm{now}}/M^4)}}~.
\end{equation}
As before, we list the values of $M$ and $m_\mathrm{now}$ for various choices of $\lambda$ in Tab. \ref{TabIEMass}.
\begin{table}[h]
\centering
\begin{tabular}{l|l|l}
\hline
$\bm{\lambda}$ & $\bm{M\,(10^{-3}\,\mathrm{eV})}$  & $\bm{m_\mathrm{now}\,\mathrm{(eV)}}$ \\
\hline
$1$ & $2.32831$  & $5.541$ \\
$2$ & $2.3283$  & $2.591$ \\
$10$ & $2.32825$  & $0.443$ \\
$100$ & $2.32817$  & $0.035$ \\
$10^3$ & $2.32807$  & $0.0028$ \\
$10^4$ & $2.32795$  & $0.000225$ \\
\hline
\end{tabular}
\caption{Same as in Tab. \ref{TabFJMass}, but for the inverse exponent potential.}
\label{TabIEMass}
\end{table}

\begin{figure}[h]
\begin{center}
\includegraphics[width=0.6\textwidth]{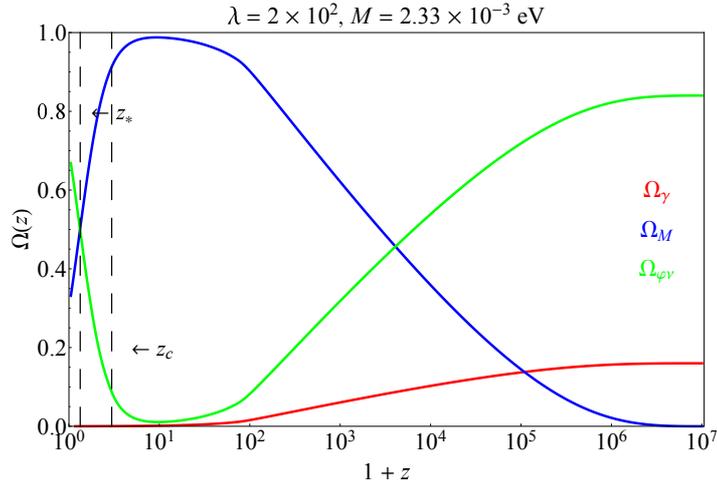}
\caption{The relative densities of photons, matter, and the $\varphi\nu$ fluid as a function of redshift for the case of the inverse exponent potential. As before, $z_\star\approx0.33$ denotes the epoch of matter-DE equality, and $z_\mathrm{c}\approx 1.97$ denotes the redshift at which the temperature crosses $T_\mathrm{c}$. The late time evolution of the $\Omega$'s are consistent with the observed evolution of the Universe.}
\label{FigOmeIE}
\end{center}
\end{figure}

\begin{figure}[h]
\begin{center}
\includegraphics[width=0.6\textwidth]{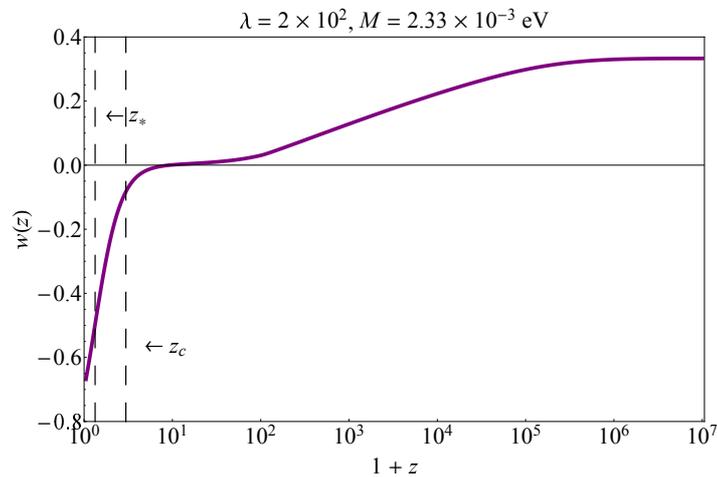}
\caption{Equation of state parameter for the entire Universe as a function of redshift for the inverse exponent potential. We see that the $w$ today is below $-1/3$, necessary for the accelerated expansion of the Universe.}
\label{FigEosZIE}
\end{center}
\end{figure}

Finally, in Figs. \ref{FigOmeIE} and \ref{FigEosZIE}, we show the relative energy densities of the various components of the Universe and  the equation of state parameter of the entire Universe. Again, we have changed the value of $\lambda$ for these two plots, choosing $\lambda=200$ for visual purposes.

The inverse exponent as a trial quintessence potential appears to be a viable candidate for the MaVaN scenario, including the late time
accelerating era of the Universe.

\section{The oscillating potential}
\label{SecOsc}

\begin{figure}[h]
\begin{center}
\includegraphics[width=0.6\textwidth]{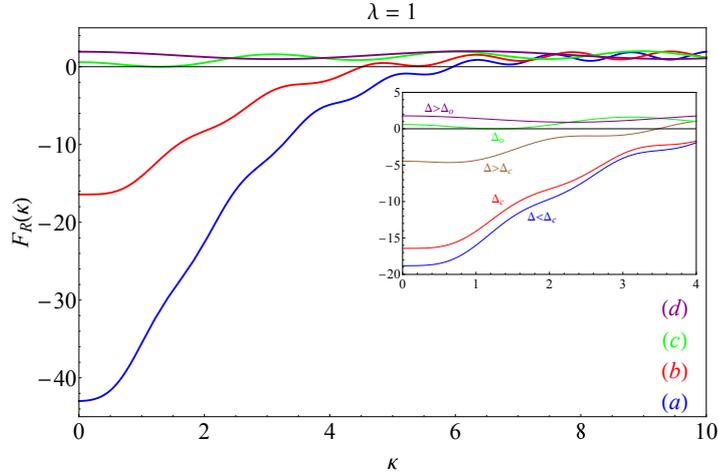}
\caption{The thermodynamic potential given by Eq. \eqref{eOscFR}, plotted for progressively increasing values of $\Delta$. The blue curve is for $\Delta<\Delta_\mathrm{c}$, the red curve corresponds to $\Delta_\mathrm{c}$ when the potential just starts to develop a minimum at $\kappa\neq 0$, the green curve is for $\Delta_\circ$, and the purple curve is for $\Delta>\Delta_\circ$. In the inset, we plot some of these features more closely. The red and green curves are the same as before, while the blue and purple curves are plotted with slightly different values of $\Delta$ for visual purposes. The brown curve denotes a stage after crossing $T_\mathrm{c}$ when the distinct nontrivial minimum can be seen.}
\label{FigFRCosSq}
\end{center}
\end{figure}

This is a potential proposed in the context of thawing and pseudo Nambu-Goldstone models (see Ch. 7 of Ref.~\cite{Amendola:2015ksp} for more details); the potential varies as $\cos^2(\varphi/M_\varphi)$, where $M_\varphi$ is as above. To utilize it for the MaVaN scenario, we shift the original ansatz \cite{Amendola:2015ksp} to the form
\begin{equation}
\label{eOscPot2}
U(\varphi)=M^4\,\left[1+\cos^2\left(\frac{\varphi}{M_\varphi}\right)\right]~.
\end{equation}
In the dimensionless notations of Eqs. \eqref{eDimLess} and \eqref{lam} the thermodynamic potential reads
\begin{equation}
\label{eOscFR}
F_R=1+\cos^2\left(\frac{\kappa\lambda}{\Delta}\right)-\frac{2N_F}{3\pi^2 \Delta^4} \mathscr{I}_{3/2}(\kappa)~,
\end{equation}
while the minimum equation is
\begin{equation}
\label{eOscGap}
\frac{\lambda \Delta^3 \pi^2}{2N_F}  \sin \left(\frac{2\kappa\lambda}{\Delta}\right) =\kappa\mathscr{I}_{1/2}(\kappa)~.
\end{equation}
One can check that at high temperatures $T>T_c$ only  a trivial solution of the minimum equation  $\kappa=0$ exists. It corresponds to the global minimum
of the thermodynamic potential Eq. \eqref{eOscFR} (see Fig. \ref{FigFRCosSq}). In this phase the fermion field is massless $m=0$, while the mass of the scalar field is found exactly to be
\begin{equation}
\label{MphiCosAbove}
  \frac{m_\varphi ^2}{2 \lambda M^2 }= \frac{T^2}{T_c^2}-1~,\quad\quad T>T_c~.
\end{equation}
At the critical temperature (found exactly from Eq. \eqref{eOscGap})
\begin{equation}
\label{TcCos}
T_c=\sqrt{\frac{12}{N_F}} \lambda M ~\leftrightarrow~
\Delta_\mathrm{c}=\frac{1}{\lambda}\sqrt{\frac{N_F}{12}}
\end{equation}
a nontrival solution of Eq. \eqref{eOscGap} corresponding to a new global minimum of the potential Eq. \eqref{eOscFR}
appears, signalling a phase transition and the fermion mass generation. To analyze the nature of this transition we expand both sides
of Eq. \eqref{eOscGap} at $T<T_c$ to the order  $\mathcal{O}(\kappa^2)$ and with the help of Eq. \eqref{I12As}, bring it  to the form
\begin{equation}
\label{kappa2}
  \kappa^2 \log \frac{\kappa}{\kappa_c}= \frac{\pi^2}{3}\left(\frac{T_c^2}{T^2} -1  \right)~,
\end{equation}
where
\begin{equation}
\label{kappaC}
  \kappa_c \equiv e^{-\mathds{C}}~,\quad\quad \mathds{C}(\lambda)= \frac{8 \lambda^4 \pi^2}{3N_F}-\log \pi +\gamma -\frac12~.
\end{equation}
We infer from Eq. \eqref{kappa2} that along with the trivial solution  $\kappa = 0$, a new root $\kappa=\kappa_c$ solves this equation at $T=T_c$.
The nature of this transition depends on the parameter $\lambda$. We define the borderline value $\lambda_*=0.59$ (for $N_F=3$) such that $\mathds{C}(\lambda_*)=0$, i.e., $\kappa_c(\lambda_*)=1$. It follows from the above equations that for $\lambda < \lambda_*$, we have $\kappa_c>1$, and so the order parameter experiences a considerable discontinuity at the critical point, indicating that the transition is strongly of the first order.\\
In this work we are particularly interested in the choice
\begin{equation}
\label{lam1}
  \lambda=1~,
\end{equation}
such that $M_\varphi=M$. Fixing a value of $\lambda$ leaves $M$ to be the only adjustable parameter.
In this case $\mathds{C}(1)=7.7$
and
\begin{equation}
\label{kappaC1}
  \kappa_c = 4.5 \times 10^{-4}~,
\end{equation}
so the transition is only very weakly of the first order, i.e., almost continuous.

Before presenting more results, let us first point out several qualitative differences between the $\varphi \nu$ model with oscillating potential and
the cases of the RP \cite{Chitov:2009ph} or the above exponential potentials. In the three latter cases, the thermodynamic potential of the $\varphi \nu$ model develops a nontrivial global minimum starting at arbitrary high temperatures, i.e., the fermions get an effective mass from the start. The system evolves with the cooling of the Universe (time) through a metastable phase (local minimum) to the critical point of instability,  where the minimum disappears and the model must reach the new global minimum  at $\varphi \to \infty$ via a discontinuous first order transition. After the critical point is crossed, the equilibrium methods are inapplicable, and the model must be analyzed in the framework of dynamics or kinetics.  On the contrary, the model coupled to oscillating potential develops a trivial global minimum at $\varphi =0$, i.e., fermions stay massless all the way until the system cools down to $T=T_c$. At $T_c$ the point $\varphi =0$ becomes a maximum, while a new nontrivial minimum which appears at $\varphi =\varphi_c$ (or at $\kappa_c$ in terms of dimensionless parameters). It evolves smoothly and stays global all the way down to $T \to 0$ (or $t \to \infty$), allowing us to use the equilibrium approach to deal with this case at all temperatures. The system stays locked near that minimum even after the thermodynamic potential becomes positive, i.e., the pressure becomes negative. The latter case, according to the standard thermodynamic wisdom \cite{LandauStat1}, signals metastability, even if the minimum is global.

In Fig. \ref{FigFRCosSq} we plot the thermodynamic potential $F_R(\kappa)$ at several temperatures $T>T_c$ and $T<T_c$ during the evolution
(we set $\lambda=1$).
The point of metastability $T_\circ <T_c$ where the potential at the global minimum vanishes was found numerically,
\begin{equation}
\label{To}
  \Delta_\circ =0.95~\longmapsto   T_\circ =1.05 M, ~~m_\circ =1.36M~.
\end{equation}

The subleading logarithmic corrections \eqref{kappa2} modify the behavior of the order parameter. In the extremely narrow vicinity of transition
$T_c-T \ll T_c$ and $m-m_c \ll m_c$ one finds $m-m_c \propto T_c-T$, but this is hardly practically important because of Eq. \eqref{kappaC1} (the fluctuations in the immediate vicinity of $T_c$ invalidate the mean-field predictions anyway). Furthermore in the ordered phase near $T_c$, we find the following approximate behavior,
\begin{equation}
\label{mBelow}
  \frac{m}{T_c} \approx \frac{\sqrt{\tau}}{ \log^{1/2} \left( \sqrt{\tau} /\kappa_c \right)}~,\quad\quad
  \tau \equiv \frac{2 \pi^2}{3} \left( 1-\frac{T}{T_c} \right)~,\quad\quad T  \lesssim T_c~.
\end{equation}
This is like the order parameter for a second-order transition, but modified by multiplicative logarithmic corrections.  Deep in the ordered phase at $T \ll T_c$  the mass is determined by the minimum of the scalar potential $U(\varphi)$,
\begin{equation}
\label{mT0}
  m \approx \frac{\pi M}{2 \lambda}~, \quad\quad T\ll T_c~,
\end{equation}
up to exponentially small fermionic corrections.

\begin{figure}[h]
\begin{center}
\includegraphics[width=0.6\textwidth]{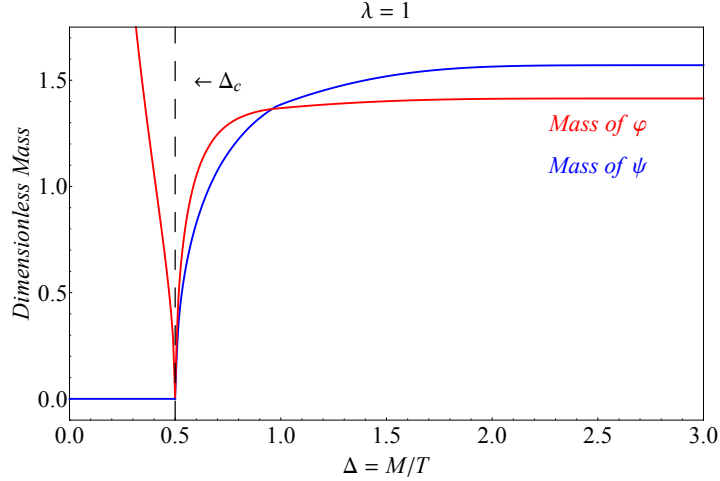}
\caption{Temperature dependence of the fermion and scalar masses for the oscillating  potential with $\lambda=1$.}
\label{FigMassesOsc}
\end{center}
\end{figure}

The mass of the scalar field near $T_c$ behaves as
\begin{equation}
\label{MphiCosBelow}
  \frac{m_\varphi ^2}{2 \lambda M^2 } \approx 2 \left( \frac{T^2}{T_c^2}-1 \right)+ \frac{3}{\pi^2} \frac{m^2}{T_c^2}~, \quad\quad T  \lesssim T_c~.
\end{equation}
Note that it has some (small, cf. Eq. \eqref{kappaC1}) residual value at $T \to T_c^-$, which is consistent with the fact that contrary to other potentials considered, $T_c$ is not an inflection point of the free energy for the case of oscillating potential.
Deep in the ordered phase, this can be written as
\begin{equation}
\label{MphiAsym}
  m_\varphi \approx \sqrt{2}  \lambda M~, \quad\quad T  \ll T_c~.
\end{equation}
The numerical solution of the equations for $m$ and $m_\varphi$ yields the results plotted in Fig. \ref{FigMassesOsc}.

\begin{figure}[h]
\begin{center}
\includegraphics[width=0.6\textwidth]{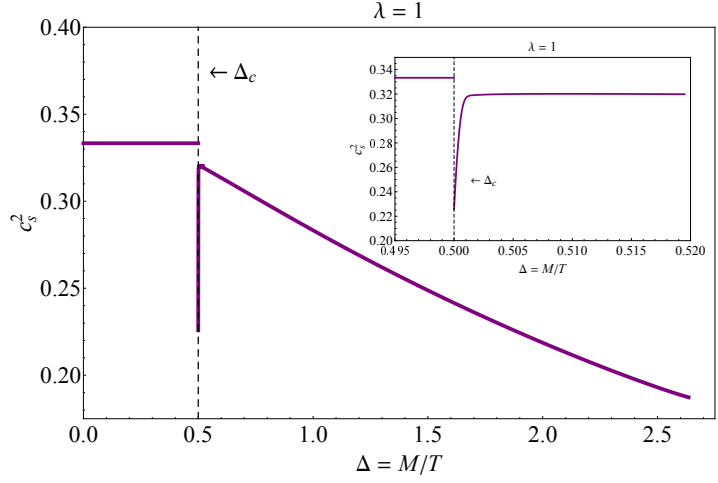}
\caption{Temperature dependence of the speed of sound $c_s^2$ in the $\varphi\nu$ fluid for the oscillating potential with $\lambda=1$. At $T>T_c$ ($\Delta<\Delta_\mathrm{c}$) $c_s^2=-1/3$.}
\label{FigCs2Osc}
\end{center}
\end{figure}

At $T>T_c$ when the minimum equation Eq. \eqref{eOscGap} is solved by the trivial root, the sound velocity Eq. \eqref{cs2Eq}
is easily found
\begin{equation}
\label{csCosAbove}
    c_s^2 = \frac{ \partial P/\partial \Delta}{\partial \rho/\partial \Delta}= \frac13~, \quad\quad T>T_c.
\end{equation}
To analyze the temperature dependence of the sound velocity in the ordered phase, given by Eq. \eqref{eCsSqInteg}, we use
Eq. \eqref{eOscGap} to derive
\begin{equation}
\label{DLogCos}
   \frac{d \log \kappa_m}{d \log \Delta } = \frac{3 - (2 \lambda \kappa/\Delta) \cot (2 \lambda \kappa/\Delta)}{1
   - (2 \lambda \kappa/\Delta)\cot (2 \lambda \kappa/\Delta)
    +d \log \mathscr{I}_{1/2}(\kappa)/d \log \kappa} \Bigg\vert_{\kappa=\kappa_m}~.
\end{equation}
At small $\kappa$ the above parameter grows as $d \log \kappa_m / d \log \Delta \propto 1/\kappa^2 \log \kappa$. However some cancellations in Eq. \eqref{eCsSqInteg} occur, and $c_s^2$ does not vanish at $T \to T_c^-$, but has a discontinuity. After some manipulations, we find a simple result for this non-zero limit,
\begin{equation}
\label{csTCmin}
  c_s^2= \frac{7}{31}=0.2258~, \quad\quad T \to T_c^-~.
\end{equation}
Using our formulas for the speed of sound deep in the ordered phase, one can recover the result of the classical ideal gas,
\begin{equation}
\label{csIgas}
  c_s^2 \approx \frac{T}{m}~,\quad\quad T \ll T_c~.
\end{equation}
which can be also written (see Eq. \eqref{mT0}) as
\begin{equation}
\label{csIgas2}
  c_s^2 \approx \frac{2 \lambda T}{\pi M}~,\quad\quad T \ll T_c~.
\end{equation}
Since  $c_s^2>0$ at all temperatures, the model is stable with respect to the density fluctuations in the massless and massive phases,
including the metastable temperature range $T< T_\circ <T_c$, where $w_{\varphi \nu}<0$.
The numerical results for the speed of sound are shown in Fig.~\ref{FigCs2Osc}.

To apply the oscillating potential for our Universe we need the current DE density to agree with the observed value  $\rho^\mathrm{now}_{\varphi\nu} \sim \mathcal{O}(M^4)$ with $M \sim 10^{-3}$ eV. We identify  the current temperature of the Universe with the cosmic background radiation temperature
\begin{equation}
\label{Tnow}
  T_{\mathrm{now}} = 2.275~K=2.4 \times 10^{-4} ~\mathrm{eV}.
\end{equation}
The critical temperature $T_c=2 M$, given by Eq. \eqref{TcCos} with $\lambda=1$ and $N_F=3$, and thus we readily establish that the present Universe is in the regime $T \ll T_c$. As was discussed above, in this regime the fermionic contribution is exponentially small and the field settles in the minimum $\varphi= \pi M/2 \lambda$, yielding
\begin{equation}
\label{eOscFRToday}
F(\varphi)|_\mathrm{today} \simeq F_R(\pi M/2 \lambda) \simeq M^4~.
\end{equation}
Thus, at the present time the energy density
\begin{equation}
\label{eOscRhoNow}
\rho^\mathrm{now}_{\varphi\nu}\simeq M^4
\end{equation}
acts essentially as a cosmological constant with
\begin{equation}\label{eOscPNow}
P^\mathrm{now}_{\varphi\nu}\simeq -M^4\quad \mathrm{and} \quad w_{\varphi\nu}=-1~,
\end{equation}
and the mass parameter
\begin{equation}\label{eOscMSc}
M=2.31\times 10^{-3}\,\mathrm{eV}~.
\end{equation}
The present mass of neutrinos is obtained to be
\begin{equation}
\label{e232}
m_\mathrm{now} = \varphi_\mathrm{now} \approx \frac{\pi M}{2 \lambda }=3.6 \times 10^{-3}\,\mathrm{eV}~.
\end{equation}
Note that without exponentially small fermionic contributions, the global minimum of the scalar field $\varphi= \pi M/2 \lambda$ would become
just one of the infinitely degenerate minima of the oscillating potential Eq. \eqref{eOscPot2}.

\begin{figure}[h]
\begin{center}
\includegraphics[width=0.6\textwidth]{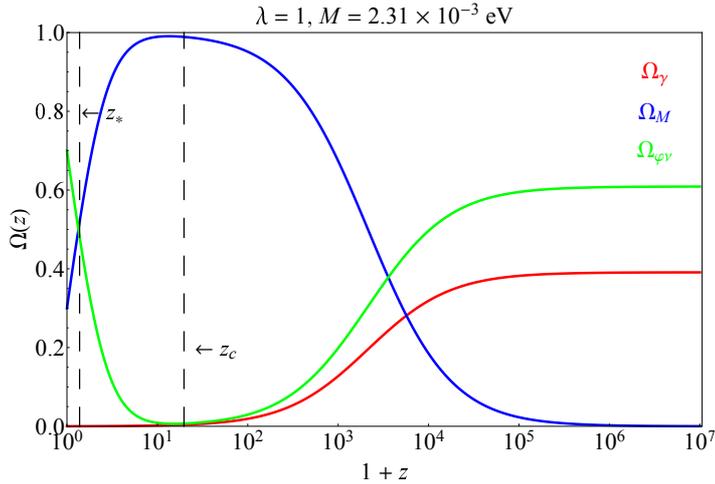}
\caption{The relative densities of photons, matter, and the $\varphi\nu$ fluid as a function of redshift for the case of the oscillating potential, and as before, $z_\star \approx 0.38$ and $z_\mathrm{c} \approx 18.3$ denote respectively the epoch of matter-DE equality, and the time at which the temperature falls below $T_\mathrm{c}$. As for the case of the inverse exponent potential, the late time evolution of the $\Omega$'s are consistent with the observed evolution of the Universe.}
\label{FigOmeOsc}
\end{center}
\end{figure}

\begin{figure}[h]
\begin{center}
\includegraphics[width=0.6\textwidth]{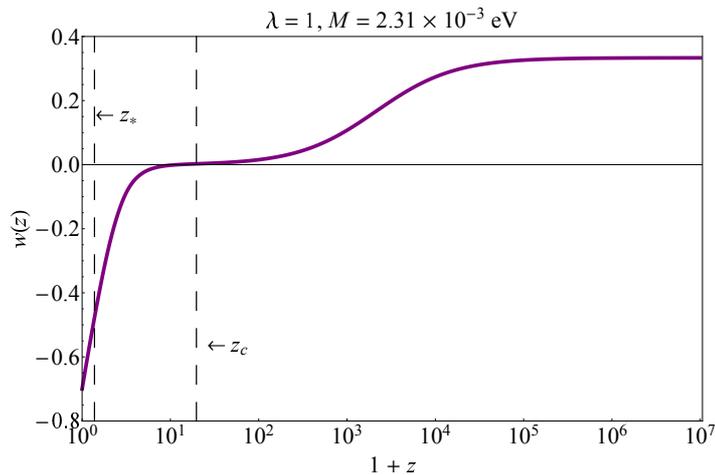}
\caption{Equation of state parameter for the entire Universe as a function of redshift for the oscillating potential; again $w$ at the present time is below $-1/3$, leading to the accelerated expansion of the Universe.}
\label{FigEosZOsc}
\end{center}
\end{figure}

As for the potentials above, we plot the relative energy densities and the equation of state parameter of the entire Universe in Figs. \ref{FigOmeOsc} and \ref{FigEosZOsc}, respectively. With the parameters chosen, the critical point $T_c =4.62 \times 10^{-3}\,\mathrm{eV}$  corresponds to
the redshift
\begin{equation}
\label{zc}
  1+z_\mathrm{c}= \frac{T_c}{T_{\mathrm{now}}} \approx 19.3~.
\end{equation}

These plots along with other results of this subsection confirm that oscillating potential in Eq. \eqref{eOscPot2} is a consistent choice for the MaVaN scenario.

\section{Conclusions and future directions}
\label{secConcl}

This paper extends the earlier work of Chitov et al. \cite{Chitov:2009ph}  by exploring the MaVaN model of Fardon, Nelson, and Weiner \citep{Fardon:2003eh}  for
more quintessence potentials. The analysis is performed in the framework of thermal quantum field theory to derive a self consistent formalism for the generation of neutrino masses by starting with massless Dirac fermions coupled to scalar field DE.
It is shown for several potentials, in line with the results \cite{Chitov:2009ph} for the Ratra-Peebles potential, that the mass equation has a nontrivial equilibrium solution in certain temperature ranges (phases). The key results of the MaVaN model as presented here are: it is able to self-consistently generate neutrino masses and reproduce the observed late-time evolution of the Universe ($w \to -1$, $U(\varphi) \sim M^4 \sim \Lambda$),
asymptotically close to that of the $\Lambda$CDM model,  which is the Standard Model of cosmology\footnote{This is not always the case. The decoupled quintessence (see, e.g., several examples in Refs.~\cite{Amendola:2015ksp,Avsajanishvili:2017zoj}) or even coupled MaVaN models \cite{Kisslinger:2019} can result in solutions which are quite far from the asymptotically approached cosmological term.}.

In this work, we have looked at the Ferreira-Joyce potential \eqref{eFJPot}, the inverse exponent potential \eqref{eIEPot}, and the oscillating potential \eqref{eOscPot2}. For the first two potentials the temporal evolution of the solution of the mass equation is qualitatively similar to that for the Ratra-Peebles potential (see Appendix \ref{eAppRP}) -- it grows smoothly from 0 to the critical value at the point of the first-order phase transition.
The Universe undergoes the following transition: the former minimum of the thermodynamic potential at a finite value of the quintessence field
becomes an unstable inflection point, and the Universe evolves towards the doomsday vacuum at $\varphi=\infty$ ($\Lambda$-term). The thermal (temporal) evolution of the Universe around the stable minimum before the transition point is analyzed by the methods of equilibrium thermal theory, while after
the transition the dynamics is described by the equations of motion. In both cases the $\varphi \nu$ model is coupled to the Friedmann equations.

The Ferreira-Joyce potential can be ruled out vis-\`{a}-vis the MaVaN model based on the late time cosmological evolution of the $\varphi\nu$ fluid. We see from Eq. \eqref{eFJRhoAfter} that after the phase transition, this fluid behaves like a pressureless matter component with an equation of state parameter $w_{\varphi\nu}=0$. This is also clearly seen in Figs. \ref{FigOmeFJ} and \ref{FigEosZFJ}. Even though the mechanism can generate acceptable neutrino masses at the present time for a range of the parameter $\lambda$ (see Tab. \ref{TabFJMass}), it cannot lead to the currently observed accelerated expansion of the Universe. This however suggests that the Ferreira-Joyce potential can be a model describing DM in the context of a MaVaN like scenario. It is known, however, that DM exists in the Universe at epochs much earlier than recombination \cite{Sarkar:2014bca}, and thus for the MaVaN mechanism with the Ferreira-Joyce potential to be a viable model for DM, the critical temperature $T_c$ has to be higher than that at recombination. The mass scales $M$ and $M_\varphi$ in such a case need to be very different from those discussed in Sec. \ref{secFJDynAft}, and there might be more features and additional degrees of freedom needed in the model; we defer further investigations into this to a future work. On the contrary, the inverse exponent potential Eq. \eqref{eIEPot} passes these hurdles and is qualified as a viable DE potential for the consistent MaVaN scenario, along with the RP potential from the earlier work \cite{Chitov:2009ph}.

The oscillating potential in Eq. \eqref{eOscPot2} turns out to be another consistent candidate for the MaVaN scenario. The latter, however has
a qualitatively different behavior compared to the case of the exponential or the RP potentials:
\begin{itemize}
\item Contrary to other potentials which have two adjustable parameters each, the periodic potential is able to reproduce the observable data with a single (mass) parameter.
\item In the range of parameters we are interested in, this potential results in a virtually continuous phase transition of the Universe at the critical point.
\item Neutrinos are massless in the high-temperature phase until quite late times $z_c \simeq 18$.
\item After the phase transition the Universe does not lose stability, but smoothly evolves into a new global minimum, which remains stable at all times $t > t_c$.
\item Sufficiently far away from the transition point $T \ll T_c$, this minimum becomes infinitely degenerate, up to exponentially small fermionic corrections, suggesting then that quantum fluctuations could trigger tunneling to other minima, leading to finite jumps of the neutrino mass.
\end{itemize}

The presence of the Yukawa interaction in the MaVaN model opens the possibility of the scalar dynamics being affected by the generation of effective $\varphi^n$ interactions for $n>2$.
In appendix \ref{eAppLoop}, we study that the contribution of such interaction towards the effective potential, and see that it does not destroy the dynamics of the scalar field in the regime important to the generation of the neutrino mass in a way consistent with the MaVaN scenario.

There is also another important difference in the MaVaN scenarios for these potentials. The instability after the critical point of the strong first-order transition predicts strong growing fluctuations of the neutrino density (neutrino clumps or clustering). The growing neutrino fluctuations
have been studied in the literature quite actively \cite{Afshordi:2005ym, Kaplinghat:2006jk, Valiviita:2008iv, Jackson:2009mz, Sawicki:2012re, Gavela:2009cy, Baldi:2011es, Ayaita:2011ay, Ayaita:2014una, Casas:2016duf, Brookfield:2006, Brookfield:2008}.
In particular, in Ref. \cite{Casas:2016duf}, the authors had shown that for neutrinos that are light currently ($\sim\mathcal{O}(0.01\,\mathrm{eV})$), the neutrino clusters form and dissolve, without leaving any imprints on the evolution history of the Universe. If the neutrino masses are much larger than this, neutrino clumps form at a much faster rate, leading the evolution of the Universe to deviate from that observed at the current time. The gravitational potential of these neutrino overdensities can impact the low $\ell$ modes of the cosmic microwave background (CMB) via the Integrated Sachs-Wolfe (ISW) effect \cite{Ayaita:2014una, Pettorino:2010bv}, and also the power spectrum of large scale structures \cite{Baldi:2011es}. The gravitational potential of these neutrino clusters can also affect the weak lensing spectrum \cite{Baldi:2011es, Ayaita:2011ay} of CMB photons. The excess lensing power reported by the Planck collaboration \cite{Aghanim:2018eyx} (see their Fig. 3) might be a signature of these clusters of neutrinos; however it is still unclear whether systematic effects in the Planck data is responsible for this excess.
Also, to the best of our knowledge, there are no unequivocal experimental data which confirm or deny the existence of these clusters.

The above consequences had been worked out for the \textit{growing neutrino quintessence} (GNQ) model \cite{Wetterich:2007kr, Ayaita:2014una}, where the coupling between neutrinos and the quintessence scalar field is different from what we have considered here. Such a model does not exhibit a first order thermodynamic phase transition as in our work (for the Ratra-Peebles, Ferreira-Joyce, and inverse exponent potentials), and as far as we know, the growth of neutrino clusters has not been studied in models that exhibit a first-order phase transition.
However, certain phenomenological features of the MaVaN scenario we have worked with are qualitatively similar to those of the GNQ model. In the latter, as the mass of the neutrinos grows with the evolution of the Universe, and they become non-relativistic, the Universe transitions to a vacuum dominated evolution (see also Ref. \cite{MohseniSadjadi:2017pne} for a different model with similar conclusions), and the neutrino overdensities start forming. It is conceivable that for the Ratra-Peebles, Ferreira-Joyce, and inverse exponent potentials, similar neutrino overdensities can form once the $\varphi\nu$ fluid transitions into the unstable phase.
The first-order phase transition associated with the Ratra-Peebles, Ferreira-Joyce, and inverse exponent potentials should also be accompanied by a release of latent heat $\Delta L=F_R(\kappa_c)M^4$ with an energy density similar to that of the DE density at the present time (within a few orders of magnitude; see Figs. \ref{FigFRFJ} and \ref{FigIEFR}), and a consequent release of entropy $\Delta S=\Delta L/T_c$. 
A microscopic analysis of the phase transition process, as well as a detailed analysis of the evolution of the perturbations in the unstable regime of the $\varphi\nu$ fluid, is needed to understand both the quantitative details of the formation of such overdensities and the effects and signatures of the release of latent heat on the evolution of the Universe.
It was also recently proposed \cite{Sakstein:2019fmf} that in addition to the coincidence problem, the Hubble tension \cite{Verde:2019ivm, Rameez:2019wdt, deJaeger:2020zpb} can also be resolved if a scalar DE interacts with neutrinos and receives an injection of energy when the masses of the neutrinos cross increase sufficiently so that they become non-relativistic.
A more detailed numerical study will throw light on whether a similar resolution can be achieved in the MaVaN scenario as studied in the present work.

The periodic potential studied in our work has no instabilities of the $\varphi\nu$ fluid, and we therefore do not expect the formation of neutrino clustering in this scenario.
The results presented here for \tcb{this} potential (one can also search for other potentials leading to a qualitatively similar behavior) open an interesting possibility to explore the MaVaN scenarios of two kinds -- one with the instabilities and neutrino clustering, and the second one without these two phenomena.
More work, especially on the numerical front, needs to be done to fully explore the observable effects of these types of MaVaN models on the expansion history of the Universe.

The models considered here are toy models of a single fermionic species coupled to the scalar field with different potentials. To get more realistic, we need to incorporate mass-varying neutrinos of different flavors in the Lagrangian of the Standard Model. A very preliminary work in this direction was
done earlier by one of us in Ref.~\cite{Chitov:2011bq}. For each flavor of neutrinos the Lagrangian consists of the left- and
right-handed components of the Dirac field $\psi$ as $\mathcal{L}_{D}= m_D \bar{\psi}_R \psi_L~+~h.c.$, with the Dirac mass arising from the Yukawa coupling to the neutral Higgs scalar, i.e., $m_D =g_D \langle H \rangle$. In addition one can assume the existence of the right-handed
Majorana neutrinos with mass $m_R$ and  a mass term,
\begin{equation}
\label{LMaj}
    \mathcal{L}_{M}= \frac12 m_R \bar{\psi}_R^c \psi_R~+ h.c.~,
\end{equation}
where the charge conjugation is defined as $\psi_R^c =- \imath \gamma^2 \psi_R^*$. In the spirit of the MaVaN scenario, it was proposed \cite{Chitov:2011bq} that the origin of the Majorana neutrino mass is due to the coupling to the DE  (quintessence), i.e., $m_R =g \langle \varphi \rangle$. This proposal was first put forward by Shaposhnikov and collaborators \cite{Asaka:2005pn,Asaka:2005an} in the framework of the so-called $\nu$-Minimal Standard Model ($\nu$MSM) extension. The difference is that in the $\nu$MSM extension the right-handed neutrinos get their masses through the coupling to the inflaton field, and not the quintessence. The well-known seesaw mechanism \cite{Petcov:1987} for the Dirac-Majorana action $\mathcal{L}_{D}+\mathcal{L}_{M}$ leads to two eigen-masses of Majorana fermions for each generation of neutrino. The light particle is naturally identified with conventional (active) neutrino, while the heavy particle with a large mass eigen-value is a candidate for the DM particle (sterile neutrino).

It is an important direction for future work to extend the preliminary results for the RP potential, as obtained in Ref.~\cite{Chitov:2011bq}, to the other DE candidates,
and, more importantly, to carry out detailed analysis of such MaVaN extensions of the Standard Model during the whole thermal history of the Universe.
The problem of abundance, masses and the life-time of the DM particles would require more advanced quantum-field theoretic methods to apply \cite{Bezrukov:2007ep,Shaposhnikov:2006xi}.

It has been pointed out many times in the literature that too many scalar fields are not very natural. There have been efforts to relate, e.g., the Higgs field to inflaton \cite{Bezrukov:2007ep,Shaposhnikov:2006xi}.
It is quite plausible that the inflaton and quintessence represent the same physical field analyzed at the different regimes of the evolution of the Universe \cite{Dimopoulos:2002}. In this context it is very tempting to advance the present MaVaN theory back in time in the effort to incorporate the inflation.

\appendix

\section{The Ratra-Peebles potential}
\label{eAppRP}
\begin{figure}[h]
\begin{center}
\includegraphics[width=0.6\textwidth]{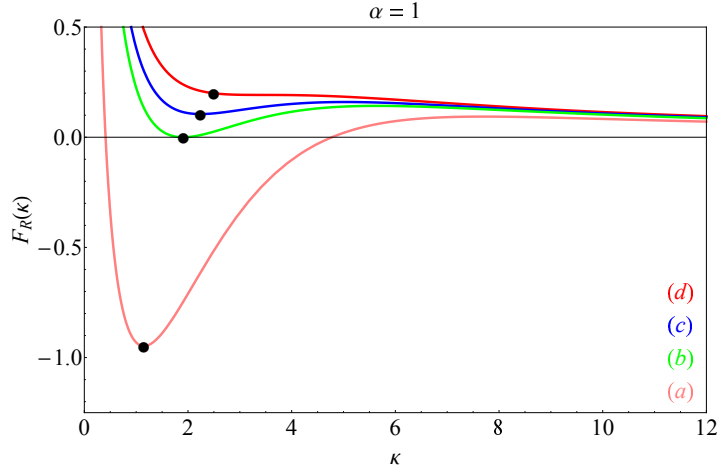}
\caption{Temporal evolution (from (a) to (d)) of the thermodynamic potential. The stable solutions of Eq. \eqref{eGapRP} are indicated by big filled dots.}
\label{FigFRRP}
\end{center}
\end{figure}

In this section, we briefly review the application of the FNW model to the Ratra-Peebles potential, due to CANK, with a few changes in notation. This potential corresponds to \cite{Ratra:1987rm},
\begin{equation}\label{eRPPot}
U(\varphi)=\frac{M^{\alpha+4}}{\varphi^\alpha}
\end{equation}
for some mass scale $M$, and $\alpha>0$. In terms of dimensionless quantities in Eq. \eqref{eDimLess},
the mass equation \eqref{eGap} becomes,
\begin{equation}\label{eGapRP}
\frac{\alpha\pi^2}{2N_F}g^\alpha\Delta^{\alpha+4}= \kappa^{\alpha+2}\mathscr{I}_{1/2}(\kappa)~.
\end{equation}
The mass of the fermion will be $T\kappa_m$, where $\kappa_m$ solves Eq. \eqref{eGapRP}; this solution is the minimum of the thermodynamic potential,
\begin{equation}\label{eFRRP}
F_R=g^\alpha\left(\frac{\Delta}{\kappa}\right)^\alpha-\frac{2N_F}{3\pi^2\Delta^4}\mathscr{I}_{3/2}(\kappa)~.
\end{equation}
In this appendix, we follow Ref.~\cite{Chitov:2009ph} in setting the Yukawa coupling $g$ to unity; as mentioned above, this is equivalent to the rescaling $g\varphi\mapsto\varphi$.
This potential is plotted in Fig. \ref{FigFRRP} for various values of $\Delta$. With appropriate approximations, the critical point is obtained to be,
\begin{equation}\label{eCritPtRP}
\kappa_\mathrm{c}\equiv\nu\approx\frac{5}{2}+\alpha,\quad \Delta_\mathrm{c}=\left(\frac{\sqrt{2}N_F}{\alpha\pi^{3/2}}\nu^\nu e^{-\nu}\right)^{1/(\alpha+4)}~.
\end{equation}

After the phase transition, the equilibrium methods no longer apply and we have to go beyond the saddle point approximation and solve the full equation of motion,
\begin{equation}\label{eEOMRP}
\ddot{\varphi}+3H\dot{\varphi}=-\frac{\pa U}{\pa\varphi}-\rho_{s,\mathrm{c}}\left(\frac{a_\mathrm{c}}{a}\right)^3~,
\end{equation}
where,
\begin{equation}\label{erhoSCritRP}
\rho_{s,\mathrm{c}}\approx\alpha M^3\left(\frac{\Delta_\mathrm{c}}{\nu}\right)^{\alpha+1}
\end{equation}
and $a$ is the scale factor of the Universe related to the redshift $z$ by $a=(1+z)^{-1}$, and the evolution of $a$ is given by the Friedmann equations.
It is seen \cite{Chitov:2009ph} that the exact numerical solution oscillates around the mean solution,
\begin{equation}\label{eMeanPhiRP}
\bar{\varphi}=\varphi_\mathrm{c}\left(\frac{1+z_\mathrm{c}}{1+z}\right)^{3/(\alpha+1)}~,\quad\quad\quad
\rho_{\bar{\varphi}}=\rho_{\varphi,\mathrm{c}}\left(\frac{1+z}{1+z_\mathrm{c}}\right)^{3\alpha/(\alpha+1)}~,
\end{equation}
where,
\begin{equation}\label{eCritPhiRP}
\varphi_\mathrm{c}\approx\frac{\nu}{\Delta_\mathrm{c}}M~,\quad\quad\quad \rho_{\varphi,\mathrm{c}}\approx\left(\frac{\Delta_\mathrm{c}}{\nu}\right)^\alpha M^4~.
\end{equation}
From the above, we can relate $M$ to other quantities,
\begin{equation}\label{eMValRP}
M=\big(\nu^\alpha\rho_{\varphi,\mathrm{now}}\big)^{(\alpha+1)/(\alpha+4)}\Delta_\mathrm{c}^{-\alpha}T_\mathrm{now}^{-3\alpha/(\alpha+4)}~.
\end{equation}
The mass of the neutrinos is calculated as $m(t)=\bar{\varphi}(t)$; in Tab. II of CANK, the values today are tabulated for various values of $\alpha$. The mass ranges from $\mathcal{O}(0.01\,\mathrm{eV})$ to $\mathcal{O}(10\,\mathrm{eV})$.

\section{Mathematical calculations for the inverse exponent potential}
\label{eAppIE}

In this section, we present some more detailed calculations for the inverse exponent potential. For that choice of $U(\varphi)$
the normalized thermodynamic potential is given by Eq. \eqref{eThermPotIE}, and the fermion mass is determined by the minimum
equation Eq. \eqref{eGapIE}.
With the large  $\kappa$ approximation we get,
\begin{equation}\label{e192}
\mathcal{I}_\Delta(\kappa)\approx \sqrt{\frac{\pi}{2}}\kappa^{7/2}\,e^{-\kappa(1+\Delta/(\kappa^2\lambda))}~.
\end{equation}
This function is maximized, for a particular $\Delta$, at $\kappa_m^\Delta$ (i.e., $\mathcal{I}'_\Delta(\kappa_m^\Delta)=0$) where,
\begin{equation}\label{e193}
\kappa_m^\Delta=\frac{7}{4}\left[1+\sqrt{1+\left(\frac{4}{7}\right)^2\frac{\Delta}{\lambda}}\right]~.
\end{equation}
We see that $\kappa_m^\Delta>7/2$ for all values of $\Delta/\lambda$. The critical value of $\Delta$, i.e., $\Delta_\mathrm{c}$, is the solution of,
\begin{equation}
\label{e194}
\frac{\pi^2}{2N_F\lambda}\Delta^5=\sqrt{\frac{\pi}{2}}(\kappa_m^\Delta)^{7/2}\,e^{-\kappa_m^\Delta(1+\Delta/((\kappa_m^\Delta)^2\lambda))}~,
\end{equation}
where $\kappa_m^\Delta$ is given by Eq. \eqref{e193}. The extremum condition $\mathcal{I}'_\Delta(\kappa_m^\Delta)=0$ gives us $-\kappa_m^\Delta\left(1+\Delta/((\kappa_m^\Delta)^2\lambda)\right)=7/2-\kappa_m^\Delta$, and thus Eq. \eqref{e194} can be rewritten as,
\begin{equation}\label{e195}
\frac{\pi^2}{2N_F\lambda}\Delta^5\sqrt{\frac{2}{\pi}}\left(\frac{2}{e}\right)^{7/2}=(2\kappa_m^\Delta)^{7/2}e^{-2\kappa_m^\Delta}~.
\end{equation}
Let us call the LHS of Eq. \eqref{e195} as $\mathscr{C}$, and the RHS as $f(2\kappa_m^\Delta)$. The function $f(\xi)$ has a maximum at $\xi_\mathrm{max}=7/2$, giving a maximum value of $f(\xi_\mathrm{max})=\left(7/2e\right)^{7/2}\approx 2.422$. Now we have seen above that $\kappa_m^\Delta>7/2$ (meaning $\xi>7$); this implies,
\begin{equation}\label{e196}
\frac{\pi^2}{2N_F\lambda}\Delta_\mathrm{c}^5\sqrt{\frac{2}{\pi}}\left(\frac{2}{e}\right)^{7/2} < \left(\frac{7}{2e}\right)^{7/2}~,
\end{equation}
giving us an upper cap on $\Delta_\mathrm{c}$,
\begin{equation}\label{e197}
\Delta_\mathrm{c} < \left[\lambda N_F \left(\frac{7}{4}\right)^{7/2} \left(\frac{2}{\pi^3}\right)^{1/2} \right]^{1/5}~.
\end{equation}
Eq. \eqref{e195}, written as $f(\xi)=\mathscr{C}$, can be solved as,
\begin{equation}\label{e198}
\xi=-\frac{7}{2}W\left(-\frac{2}{7}\mathscr{C}^{2/7}\right)~,
\end{equation}
where $W(x)$ is the \textit{Lambert W function}. We can also write Eq. \eqref{e195} as,
\begin{equation}\label{e199}
\xi-\frac{7}{2}\log\xi=-\log\mathscr{C}\equiv u~,
\end{equation}
which can be inverted to give,
\begin{equation}\label{e200}
\begin{aligned}
\xi &\simeq u\left[1+\log u\left\{\frac{7}{2u}+\left(\frac{7}{2u}\right)^2+\cdots\right\}\right] =-\log\mathscr{C}\\
&\times\left[1+\log\big(-\log\mathscr{C}\big)\left\{-\frac{7}{2\log\mathscr{C}}+\left(\frac{7}{2\log\mathscr{C}}\right)^2+\cdots\right\}\right]~.
\end{aligned}
\end{equation}
We know that,
\begin{equation}\label{e201}
\mathscr{C}=\frac{\pi^2}{2N_F\lambda}\Delta^5\sqrt{\frac{2}{\pi}}\left(\frac{2}{e}\right)^{7/2}=A\frac{\Delta^5}{\lambda}~,
\end{equation}
where $A=8\pi^{3/2}/(N_Fe^{7/2})=0.448397$ with $N_F=3$. Let us write the solution in Eq. \eqref{e200} as a function $\omega$,
\begin{equation}\label{e202}
\kappa_m^\Delta=\omega\left(\mathscr{C}\right)=\omega\left(A\frac{\Delta^5}{\lambda}\right)~.
\end{equation}
Comparing with Eq. \eqref{e193}, we get,
\begin{equation}\label{e203}
\omega^2-\frac{7}{2}\omega-\frac{\Delta}{\lambda}=0~.
\end{equation}
We expect $\lambda\gg 1$ (since $U\sim M^4$), and therefore $\mathscr{C}\ll1$ (see Eq. \eqref{e201}); and we can thus approximate,
\begin{equation}\label{e204}
\omega\left(\mathscr{C}\right)\simeq-\frac{1}{2}\log\mathscr{C}~.
\end{equation}
This tells us that $\Delta/\lambda\ll\omega,\omega^2$, and thus from Eq. \eqref{e203}, we get $\omega\simeq7/2$. From Eq. \eqref{e204}, we finally get the critical $\Delta$,
\begin{equation}\label{e205}
\Delta_\mathrm{c}\simeq\left(\frac{\lambda}{A}\right)^{1/5}\,e^{-7/5}~.
\end{equation}
Binomially expanding Eq. \eqref{e193}, and putting in the expression for $\Delta_\mathrm{c}$, we finally get,
\begin{equation}\label{e206}
\kappa_\mathrm{c}=\frac{7}{2}+\frac{2}{7}\frac{e^{-7/5}}{(\lambda^4A)^{1/5}}~.
\end{equation}
The expression for the mass of the scalar can be written,
\begin{equation}\label{e207}
\frac{m_\varphi}{M}=\left(\frac{\Delta}{\kappa_m}\right)^2\Lambda\,e^{\Delta/2\kappa_m\Lambda}\sqrt{1+\frac{2\kappa_m\Lambda}{\Delta}}~,
\end{equation}
where as before, $\kappa_m$ solves the mass equation Eq. \eqref{eGapIE}.\\

\subsection*{After the phase transition}
The scalar fermion density can is obtained from the solution for the evolution equation for $\varphi$ after the transition,
\begin{equation}\label{e208}
\frac{\rho_s}{M^3}=\left(\frac{\Delta}{\kappa}\right)^2\frac{e^{\Delta/\kappa\lambda}}{\lambda}~.
\end{equation}
From Eqs. \eqref{e205} and \eqref{e206}, we get,
\begin{equation}\label{e209}
\frac{\Delta_\mathrm{c}}{\kappa_\mathrm{c}}=\displaystyle{\frac{\lambda}{\frac{2}{7}+\frac{7}{2}e^{7/5}(a\lambda^4)^{1/5}}}~.
\end{equation}
So, the scalar density at the critical point is given by,
\begin{equation}\label{e210}
\frac{\rho_{s,\mathrm{c}}}{M^3}=\frac{\lambda}{S^2}e^{1/S}~,
\end{equation}
where $S=2/7+7e^{7/5}(a\lambda^4)^{1/5}/2$. Similarly, the critical energy density is given by,
\begin{equation}\label{e211}
\frac{\rho_{\varphi,\mathrm{c}}}{M^4}=e^{1/S}~.
\end{equation}
The neutrino mass today is thus calculated to be,
\begin{equation}\label{e212}
m_\mathrm{now}=\varphi_\mathrm{now}=\frac{M}{\lambda}\displaystyle{\frac{1}{\log(\rho_{\varphi,\mathrm{now}}/M^4)}}~,
\end{equation}
and get an estimation of $M$,
\begin{equation}\label{e214}
M=\frac{S^2e^{-1/S}\rho_{\varphi,\mathrm{now}}}{(T_\mathrm{now}\Delta_\mathrm{c})^3}\log^2\frac{\rho_{\varphi,\mathrm{now}}}{M^4}~.
\end{equation}

\section{1-loop corrections to the effective potential}
\label{eAppLoop}

In this section, we look at the contribution of bosonic and fermionic correstions at the one-loop level to the effective thermodynamic potential of the quintessence field.
From Eq.~\eqref{eFermMass}, we see that the fermion acquires an effective mass given by the value of the scalar field that minimises the effective potential $F_{\varphi\nu}(\varphi)$.
Taking these corrections into account, one can write the net thermodynamic potential as
\begin{equation}\label{ePot1Loop-1}
F^{\rm 1-loop}_{\varphi\nu}(\varphi)=F_{\varphi\nu}(\varphi)+\Delta F^F(\varphi)+\Delta F^B(\varphi),
\end{equation}
where $\Delta F^F(\varphi)$ and $\Delta F^B(\varphi)$ are respectively the 1-loop fermionic and bosonic corrections \cite{Linde:1978px,Weinberg:1996kr},
\begin{equation}\label{e1LoopTerms}
\begin{aligned}
\Delta F^F(\varphi)&=-\frac{\varphi^4}{16\pi^2}\left(\log\frac{\varphi^2}{m^2}-\frac{1}{2}\right),\\
\Delta F^B(\varphi)&=\frac{1}{64\pi^2}\left(F''_{\varphi\nu}\right)^2\left[\log\left(\frac{F''_{\varphi\nu}}{m^2_\varphi}\right)-\frac{1}{2}\right],
\end{aligned}
\end{equation}
with $m^2_\varphi$ given by Eq.~\eqref{eRenormMass}.
The renormalization condition used in deriving the corrections in Eq.~\eqref{e1LoopTerms} is that the minimum of the potential $F_{\varphi\nu}(\varphi)$ is preserved upon adding these contributions.
In terms of the dimensionless quantities in Eq.~\eqref{eDimLess}, we can write
\begin{equation}\label{ePot1Loop-2}
\begin{aligned}
F^{\rm 1-loop}_R(\kappa)&\equiv \frac{F^{\rm 1-loop}_{\varphi\nu}}{M^4}\\
&=F_R(\kappa)-\frac{\kappa^4}{16\pi^2\Delta^4}\left(\log\frac{\kappa^2}{\kappa_m^2}-\frac{1}{2}\right)+\frac{1}{64\pi^2}\left(F''_R(\kappa)\right)^2\left[\log\left(\frac{M^2\Delta^2}{m^2_\varphi}F''_R(\kappa)\right)-\frac{1}{2}\right],
\end{aligned}
\end{equation}
where
\begin{equation}\label{eOmegaDD}
F''_R(\kappa)=U''(\kappa)-\frac{2N_F}{3\pi^2\Delta^4}\mathscr{I}''_{3/2}(\kappa).
\end{equation}
It is clear from above that the bosonic corrections only contribute in the regimes where the logarithm in Eq.~\eqref{ePot1Loop-2} are real, i.e., $F''_R(\kappa)>0$ (for consistency, we consider the fermionic corrections in these regimes too); see discussion in Sec. 11.4 of Ref.~\cite{Peskin:1995ev}. Moreover, the corrections are the most accurate near the minimum of the potential $F_R$, and cannot be trusted far away from this minimum \cite{Linde:1978px, Peskin:1995ev}.

\begin{figure}
    \centering
    \begin{subfigure}[b]{0.45\textwidth}
        \includegraphics[width=\textwidth]{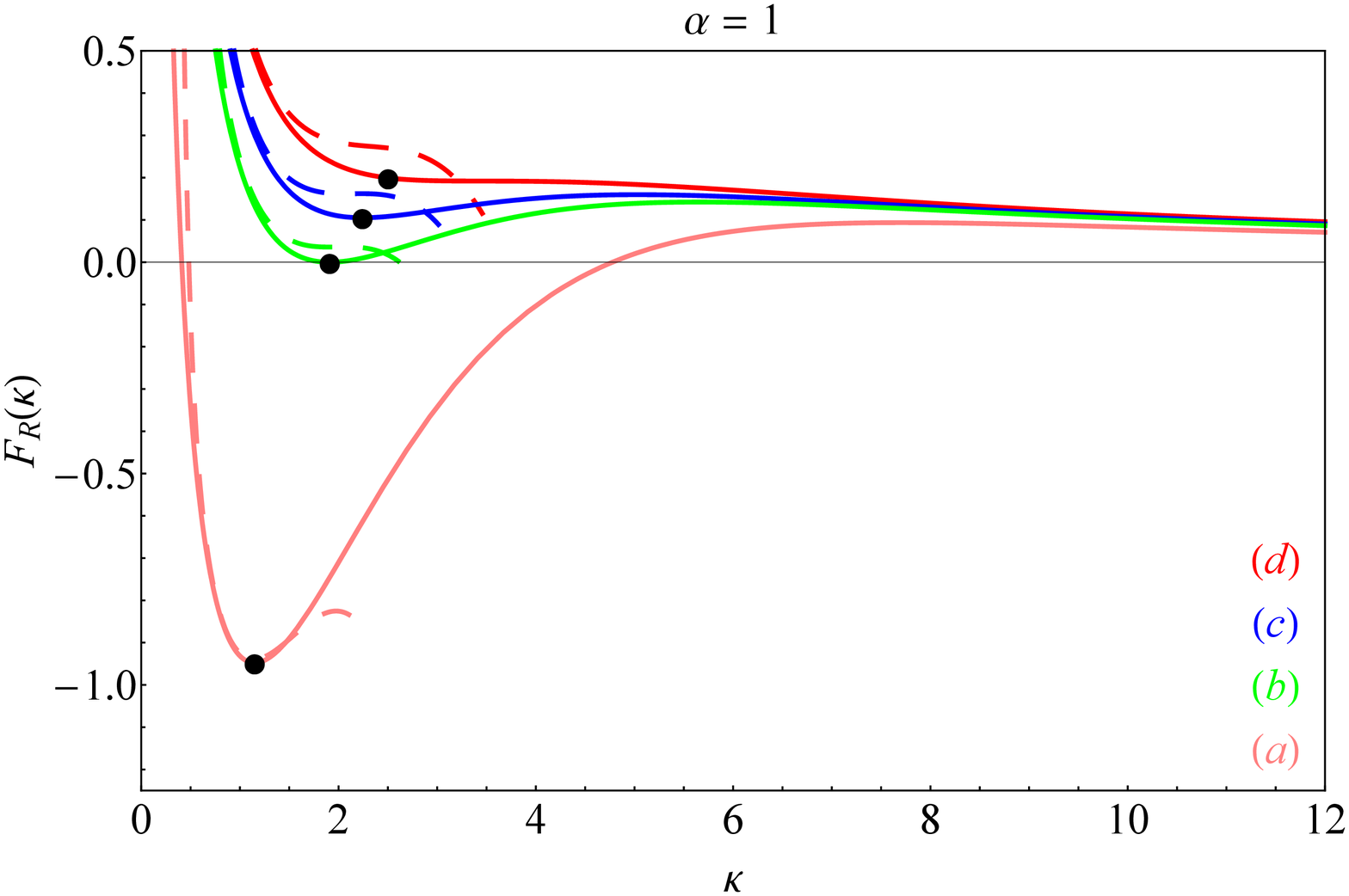}
        \caption{The RP Potential, with the same colors as in Fig.~\ref{FigFRRP}.}
        \label{FigOneLoopRP}
    \end{subfigure}
    ~
    \begin{subfigure}[b]{0.45\textwidth}
        \includegraphics[width=\textwidth]{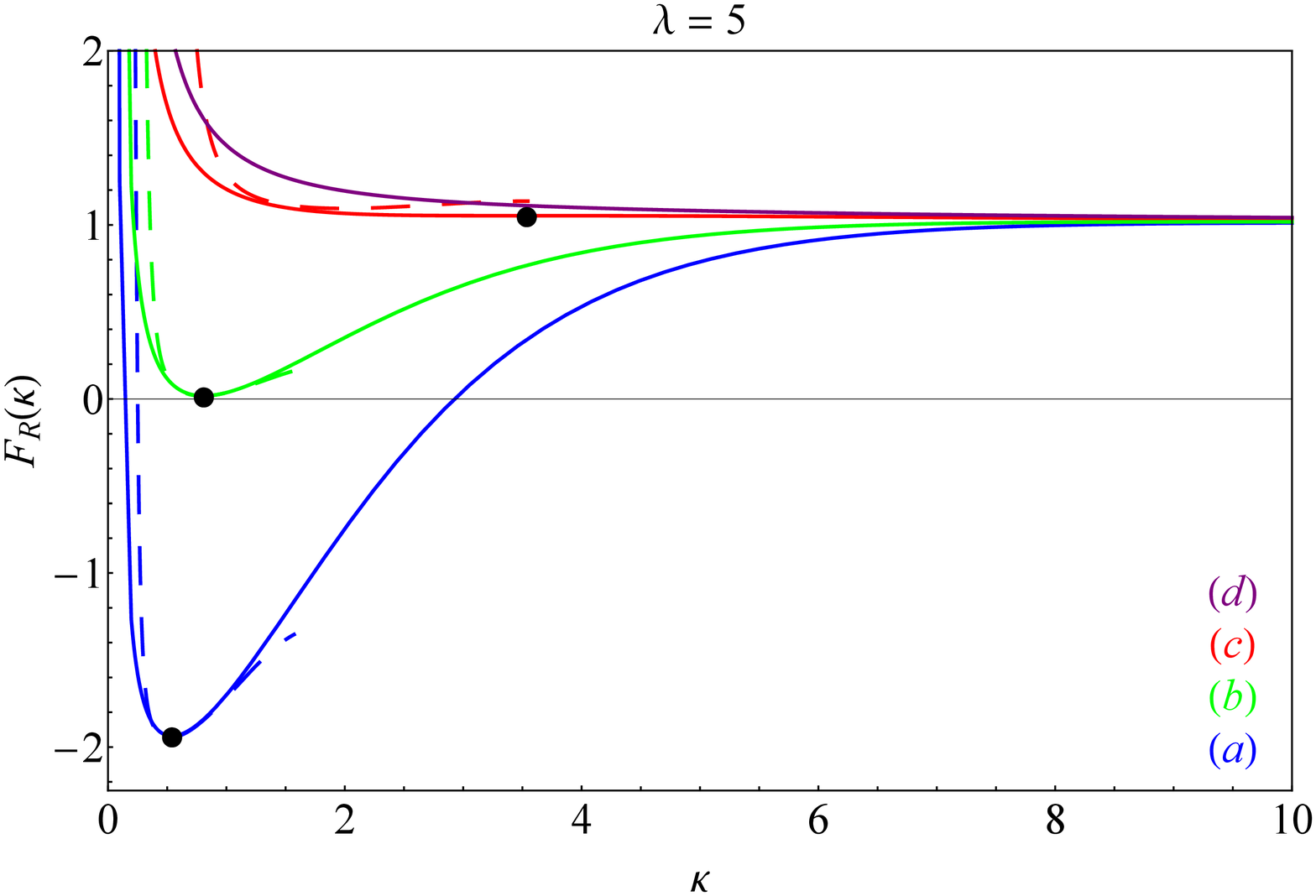}
        \caption{The IE Potential, with the same colors as in Fig.~\ref{FigIEFR}.}
        \label{FigOneLoopIE}
    \end{subfigure}
    
    \begin{subfigure}[b]{0.45\textwidth}
        \includegraphics[width=\textwidth]{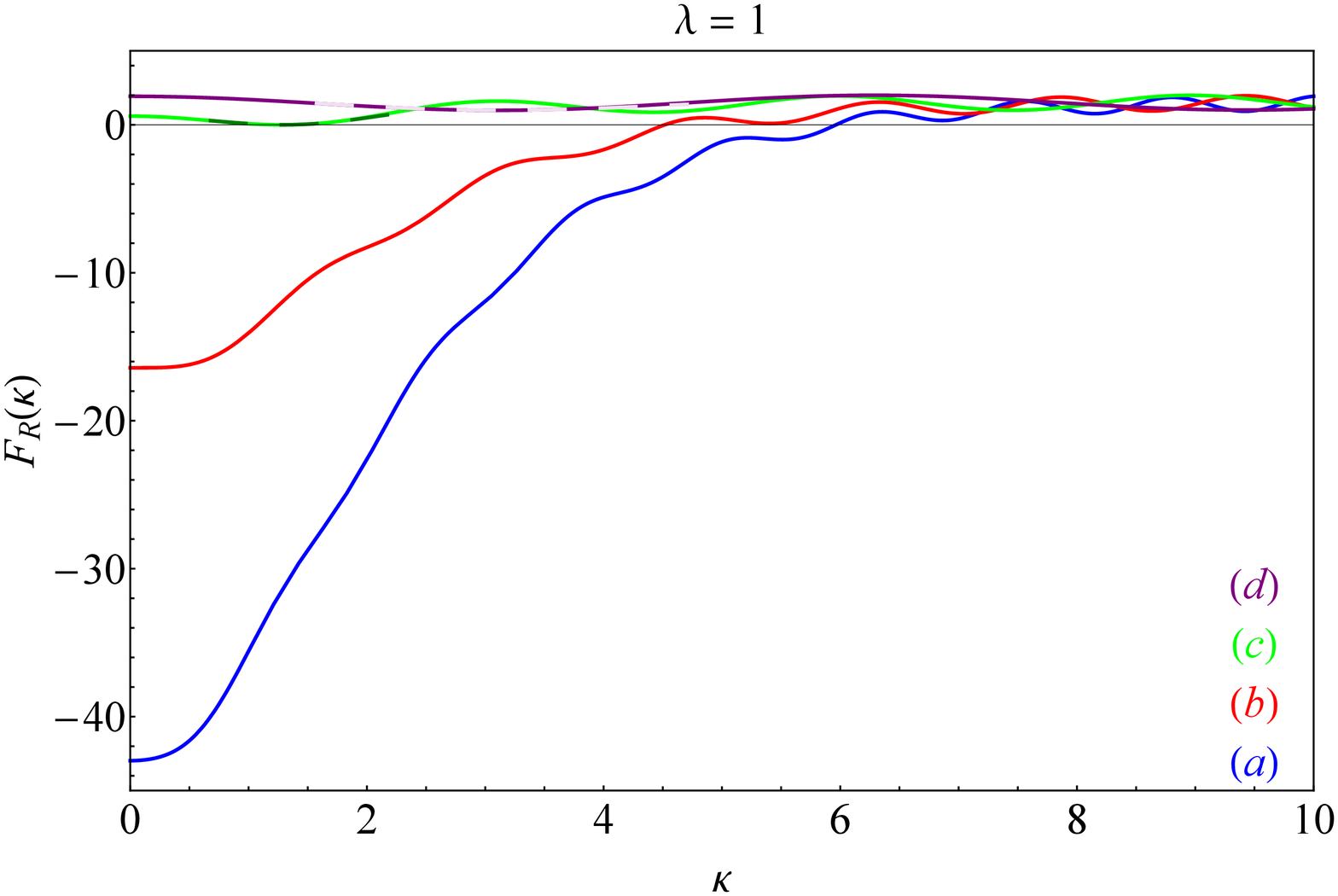}
        \caption{The Cosine Potential, with the same colors as in Fig.~\ref{FigFRCosSq}. For clarity, the correction to the solid purple line is shown as a dashed \textit{light purple} line, while that to the solid green line is shown by a dashed \textit{dark green} line.}
        \label{FigOneLoopCos}
    \end{subfigure}
    \caption{The one-loop corrected potential of Eq.~\eqref{ePot1Loop-1} (dashed) compared to the tree-level thermodynamic potential $F_{\varphi\nu}$ (solid) for the three potentials of interest.}\label{FigOneLoopCorrs}
\end{figure}

In Fig.~\ref{FigOneLoopCorrs}, we plot the 1-loop corrected potential for the three potentials of interest (we ignore the Ferreira-Joyce potential since it is not viable for generating neutrino masses in the MaVaN scenario).
It is seen that the corrections $\Delta F^F$ and $\Delta F^B$ do not significantly alter the thermodynamic potential (and hence the MaVaN dynamics) near the minimum of the thermodynamic potential $F_R$. In Figs.~\ref{FigOneLoopRP} and \ref{FigOneLoopIE}, we study these corrections in the stable and metastable regimes on the $\varphi\nu$ fluid ($T\geq T_c$), where the methods of equilibrium thermodynamics can be used to calculate the neutrino mass. As mentioned above for these two potentials, the $\varphi\nu$ fluid first-order transitions into an unstable regime where the analysis of the scalar field dynamics requires methods of non-equilibrium thermodynamics; such an analysis is beyond the scope of this work.
~\\\\
The situation for the sinusoidal potential as shown in Fig.~\ref{FigOneLoopCos} is also interesting. Since the fermion mass is zero at $T>T_c$, the one-loop corrections to the potential vanish in this regime (both $\varphi$ and $F''_{\varphi\nu}(\varphi)$ are zero at the minimum of $F_{\varphi\nu}$). After the phase transition, the corrections are non-zero, but they hardly alter the potential, as seen in Fig.~\ref{FigOneLoopCos}.
~\\\\
It is therefore seen that the fermionic and bosonic one-loop quantum corrections to the thermodynamic potential do not destroy the dynamics of the scalar field in the MaVaN scenario in any appreciable way, and thus it is a consistent mechanism to generate neutrino masses.

\acknowledgments
We thank Markus Deserno, Scott Dodelson, Yiwen Huang, Leonard Kisslinger, Arthur Kosowsky, Andrew Long, Bharat Ratra, Ira Rothstein, Lado Samushia, and Ana Paula Vizcaya Hern\'{a}ndez for many interesting and helpful discussions.
We thank the anonymous referee, the editor, and Michael Maziashvili for sugegstions and discussions regarding the one-loop correction to the potential.
T.K. and S.M. acknowledge financial support from NSF Astrophysics and Astronomy Grant (AAG) Program (Grant No. 1615940) and the Georgian National Science Foundation  GNSF (Grant No. FR/18-1462). G.Y.C. acknowledges financial support from the Laurentian University Research Fund (LURF) and the Social Sciences and Humanities Research Council (SSHRC) (Canada).
T.K. thanks ICTP for hospitality and acknowledges the support from the ICTP associate membership  program.
S.M. would also like to thank the Bruce McWilliams Graduate Fellowship for financial support.

\bibliographystyle{JHEP} \bibliography{ref}

\end{document}